\documentclass[namedreferences]{solarphysics}
\usepackage[hyperref,optionalrh,solaromanenum]{spr-sola-addons} 
\usepackage{graphicx}
\usepackage{color}
\usepackage{breakurl}          

\newcommand{\aap}{    {\it Astron. Astrophys.}}

\newcommand{\apj}{    {\it Astrophys. J.}}
\newcommand{\apjl}{   {\it Astrophys. J. Lett.}}
\newcommand{\apss}{   {\it Astrophys. Spa. Sci.}}

\newcommand{\jgr}{    {\it J. Geophys. Res.}}

\newcommand{\pasj}{   {\it Publ. Astron. Soc. Japan}}

\newcommand{\solphys}{{\it Solar Phys.}}

\newcommand{\ssr}{    {\it Space Sci. Rev.}}

\ifx \doiurl    \undefined \def
\doiurl#1{\href{http://dx.doi.org/#1}{\textsf{DOI}}}\fi \ifx \adsurl
\undefined \def
\adsurl#1{\href{http://adsabs.harvard.edu/abs/#1}{\textsf{ADS}}}\fi
\ifx \arxivurl  \undefined \def
\arxivurl#1{\href{http://arxiv.org/abs/#1}{\textsf{arXiv}}}\fi

\begin{document}
\begin{article}
\begin{opening}

\title{The 26 December 2001 Solar Eruptive Event Responsible for GLE63.
III.~CME, Shock Waves, and Energetic Particles}

\author{\inits{V.V.}\fnm{V.V.}~\lnm{Grechnev}\orcid{0000-0001-5308-6336}}
\author{V.I.~\surname{Kiselev}$^{1}$\sep
        A.M.~\surname{Uralov}$^{1}$\sep
        K.-L.~\surname{Klein}$^{2}$\sep
        A.A.~\surname{Kochanov}$^{1}$}

 \runningauthor{V.V. Grechnev \textit{et al.}}
 \runningtitle{CME and Shock Waves in the 26 December 2001 Event}

\institute{$^{1}$ Institute of Solar-Terrestrial Physics SB RAS,
                  Lermontov St.\ 126A, Irkutsk 664033, Russia;
                  email: \url{grechnev@iszf.irk.ru}
                  email: \url{valentin_kiselev@iszf.irk.ru}
                  email: \url{uralov@iszf.irk.ru}
                  email: \url{kochanov@iszf.irk.ru}\\
           $^{2}$ LESIA-UMR 8109, Observatoire de Paris, CNRS, Univ. Paris 6 \& 7, Observatoire de Meudon,
           F-92195 Meudon, France; email: \url{ludwig.klein@obspm.fr}}

\date{Received ; accepted }

\begin{abstract}
The SOL2001-12-26 moderate solar eruptive event (GOES importance
M7.1, microwaves up to 4000 sfu at 9.4 GHz, CME speed
1446\,km\,s$^{-1}$) produced strong fluxes of solar energetic
particles and ground-level enhancement of cosmic-ray intensity
(GLE63). To find a possible reason for the atypically high proton
outcome of this event, we study multi-wavelength images and dynamic
radio spectra and quantitatively reconcile the findings with each
other. An additional eruption probably occurred in the same active
region about half an hour before the main eruption. The latter
produced two blast-wave-like shocks during the impulsive phase. The
two shock waves eventually merged around the radial direction into a
single shock traced up to $25\,\mathrm{R}_\odot$ as a halo ahead of
the expanding CME body, in agreement with an interplanetary Type II
event recorded by the \textit{Radio and Plasma Wave Investigation}
(WAVES) experiment on the \textit{Wind} spacecraft. The shape and
kinematics of the halo indicate an intermediate regime of the shock
between the blast wave and bow shock at these distances. The results
show that i)~the shock wave appeared during the flare rise and could
accelerate particles earlier than usually assumed; ii)~the particle
event could be amplified by the preceding eruption, which stretched
closed structures above the developing CME, facilitated its lift-off
and escape of flare-accelerated particles, enabled a higher CME
speed and stronger shock ahead; iii)~escape of flare-accelerated
particles could be additionally facilitated by reconnection of the
flux rope, where they were trapped, with a large coronal hole;
iv)~the first eruption supplied a rich seed population accelerated
by a trailing shock wave.
\end{abstract}
\keywords{Coronal Mass Ejections; Cosmic Rays, Solar; Energetic
Particles; Flares; Radio Bursts; Waves, Shock}

\end{opening}

\section{Introduction}
\label{S-introduction}

Solar energetic particles (SEP), which are accelerated in
association with solar eruptive events, pose a hazard for
equipment and astronauts on spacecraft, and even for crew members
and passengers on aircraft in high-latitude flights because of
secondary particles produced in the Earth's atmosphere. SEPs
mainly consist of protons, $\alpha$-particles, and heavier ions.
Their energies reach hundreds of MeV and sometimes up to several
GeV. The highest-energy extremity of SEPs occasionally produces
considerable fluxes of secondary neutrons observed as ground-level
enhancements (GLE) of cosmic-ray intensity. Seventy-two GLEs have
been registered since 1942 up to the present time, mainly with
high-latitude neutron monitors (see, \textit{e.g.},
\citealp{Cliver2006, Belov2010, Nitta2012, Miroshnichenko2013} and
references therein). On average, GLEs occur once a year, but very
irregularly. GLEs avoid solar minima, while four GLEs occurred
within one week in May 1990. The rareness of GLEs hampers
understanding their origins and emphasizes the importance of
studying each solar event responsible for a GLE.

One presumable source of SEPs and GLEs is traditionally associated
with flare processes in coronal magnetic fields of active regions
exhibited in X-ray and microwave emissions. Another probable
source of SEPs is related to bow shocks driven by fast coronal
mass ejections (CMEs). In spite of the high practical importance
of SEP events, consensus has not been reached so far about the
probable contributions from the two sources in different events
and energy ranges. The main subject of the debates is related to
the origins of high-energy SEPs and especially GLEs  (see,
\textit{e.g.}, \citealp{KleinTrottet2001, Kallenrode2003,
Grechnev2008, Reames2009a, Aschwanden2012, Miroshnichenko2013} for
a review and references). Each competing concept is supported by
convincing arguments \citep{Tylka2005, Chupp2009, Vilmer2011,
Rouillard2012, Reames2013}.

The traditional view on the SEP origins is mainly based on the
hypotheses proposed in the past decades, when observational
opportunities were strongly limited relative to modern ones.
Traditional concepts considered the processes responsible for
acceleration of particles in flares and those by shock waves to be
remote and completely independent of each other. Observational
studies of the two last decades update the view on solar eruptive
phenomena step by step and establish their close association with
each other.

\cite{Zhang2001} and \cite{Temmer2008, Temmer2010} found
synchronization between the CME acceleration pulse and hard X-ray
(HXR) and microwave bursts. \cite{Qiu2007} established that the
helical component of the CME's flux rope (responsible for its
acceleration) is formed by reconnection, which caused a flare.
\cite{Miklenic2009} found a quantitative correspondence between
the reconnected magnetic flux and the rate of flare energy
release. \cite{Grechnev2011, Grechnev2013a, Grechnev2015b,
Grechnev2016} established that waves were impulsively excited by
erupting flux ropes inside developing CMEs during the rise phase
of HXR and microwave bursts and rapidly steepened into the shocks
because of a rapid falloff of the fast-mode speed. Then the shock
wave quasi-freely propagates for some time like a decelerating
blast wave and changes to the bow-shock regime later, if the
trailing CME is fast.

These results and the outlined scenario show that the traditional
contrasting of the acceleration in a flare and by a shock might be
exaggerated. Two consequences are important for the SEP
acceleration issue. First, shock waves appear much earlier than
previously assumed and can accelerate heavy particles even during
the flare. Second, a close association is expected between the
parameters of the CME, shock wave, and flare, on the one hand, and
those of a SEP event.

These circumstances indicate that both flare-related and
shock-related acceleration can be significant in SEP production,
while their roles might depend on particular conditions in
different events. Recent studies by \cite{Dierckxsens2015,
Trottet2015, Grechnev2015a} confirmed this idea and indicated
statistically increasing importance of the flare-related particle
acceleration at higher energies. The shock-related contribution
was also manifest.

In Article~I \citep{GrechnevKochanov2016} and Article~II
\citep{Grechnev2017} we started analyzing the SOL2001-12-26 event
related to an M7.1 flare with a peak time at 05:40 (all times
hereafter refer to UTC if not specified otherwise) responsible for
GLE63. Among all GLE-related flares of Solar Cycle 23, this flare
had the lowest GOES importance and longest duration, being
associated with a moderate microwave burst. Limited observations
of the flare and eruption determined incomplete knowledge of this
solar event. No soft X-ray (SXR) images or HXR data are available.
Observations with the \textit{Extreme-ultraviolet Imaging
Telescope} (EIT: \citealp{Delaboudiniere1995}), onboard the
\textit{Solar and Heliospheric Observatory} (SOHO), had a gap from
04:47 to 05:22.

Some aspects of this event look challenging. If protons and
heavier ions were accelerated in the flare concurrently with
electrons, then it is not clear why the SEP fluxes were so large.
If they were shock-accelerated, then it is not clear why the fast
CME and strong shock developed in association with a moderate
flare. It is also not clear when and where the shock wave appeared
and how it evolved.

Articles~I and II analyzed the event from microwave imaging
observations with the \textit{Siberian Solar Radio Telescope}
(SSRT: \citealp{Smolkov1986, Grechnev2003}) at 5.7\,GHz; the
\textit{Nobeyama Radioheliograph} (NoRH; \citealp{Nakajima1994})
at 17 and 34 GHz and total flux data of \textit{Nobeyama Radio
Polarimeters} (NoRP: \citealp{Nakajima1985}), and the ultraviolet
(UV) images from the \textit{Transition Region and Coronal
Explorer} (TRACE: \citealp{Handy1999}) in 1600\,\AA.

The results of Articles~I and II related to the particle event are
as follows.

\begin{enumerate}

\item GLE63 was most likely caused by the M7.1 event in Active
Region (AR) 9742 (N08\,W54). Implication of a hypothetical
concurrent far-side event is unlikely.

\item The flare was much longer than other GLE-related flares and
consisted of two parts, each of which was most likely caused by a
separate eruption.

\item The first eruption presumably occurred in AR\,9742 around
04:40 and produced ejecta. They were not observed. A related
moderate two-ribbon flare involved medium magnetic fields and
reached a GOES importance of M1.6.

\item The second eruption occurred in AR\,9742 around 05:04 and
produced a fast CME. The related main two-ribbon flare involved
strong magnetic fields associated with a sunspot and reached an
importance of M7.1.

\item An additional sharp jet-like eruption around 05:09 may have
produced a shock wave.

\end{enumerate}

Based on these results, in this article we analyze the eruptions
in this event from indirect observations. We endeavor to
reconstruct the CME and shock wave, their evolution, and to find
which circumstances could amplify the SEP outcome of this event.
Pursuing the last issue, we compare the 26 December 2001 event
with other SEP and GLE events. Invoking the recent observational
conclusions about scenarios of the CME and shock-wave development
listed in this section, we revisit this historical GLE-related
event on the basis of the modern view.

Section~\ref{S-SEP} continues the introduction and presents an
overview of the main features of the SEP event.
Section~\ref{S-eruptions} outlines the flare and reveals the
eruptions. Analyzing drifting radio bursts in a wide frequency
range, Section~\ref{S-radio bursts} reconstructs the eruptive
event, reveals the shock waves, and addresses a long-standing
issue of the relation between metric and interplanetary Type~II
events. Section~\ref{S-CME} considers the CME.
Section~\ref{S-discussion} discusses the results, evolution of the
CME and shock wave, indications of particle release, and the
possible causes of the enhanced SEP outcome of this event.
Section~\ref{S-summary} summarizes the conclusions of the study.


\section{Overview of the Particle Event}
 \label{S-SEP}

This section outlines the 26 December 2001 SEP event, whose
possible sources are in question. We list its main properties
found by different authors and comment on their conclusions. We
compare the 26 December 2001 event with other SEP and GLE events
to find their similarity, possible differences, and hints at the
probable causes of its enhanced SEP outcome. The arguments in
favor of each competing concept of the SEP origin are listed. More
information can be found in the references cited.

\subsection{Near-Earth Proton Enhancement}

\begin{figure} 
  \centerline{\includegraphics[width=0.75\textwidth]
   {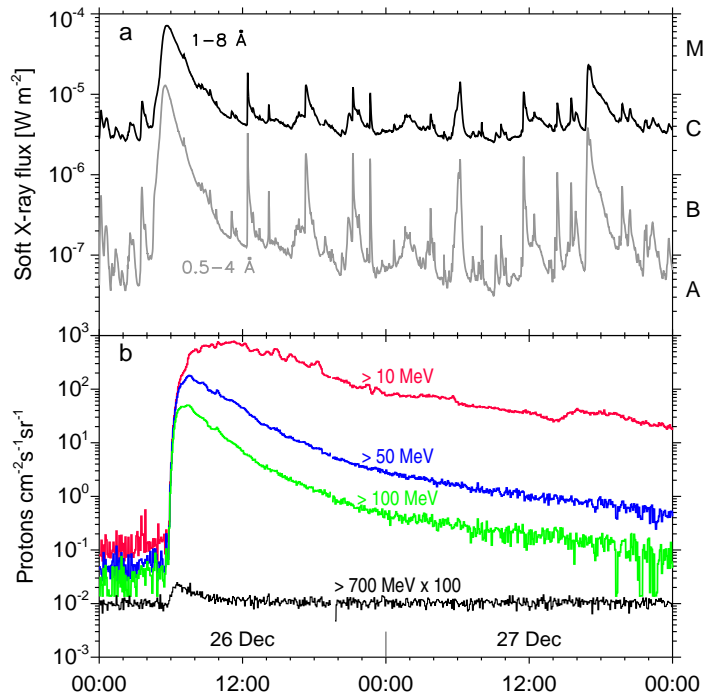}
  }
  \caption{Time profiles of the soft X-ray flux (a) and near-Earth
proton enhancement (b) recorded by three standard integral
channels of GOES-8 and the high-energy HEPAD detector ($>
700$\,MeV: black, magnified by a factor of 100).}
  \label{F-xray-protons}
  \end{figure}

SEPs are dominated by accelerated protons.
Figure~\ref{F-xray-protons} presents two-day time-profiles of the
SXR flux from the flare and the proton flux in three standard
integral channels of GOES-8. Figure~\ref{F-xray-protons}b
additionally shows the flux of high-energy protons $> 700$\,MeV
recorded by the \textit{High-Energy Proton and Alpha Detector}
(HEPAD) on GOES-8 magnified by a factor of 100.

The pre-event background does not show any elevated seed
population, as was the case before GLE33 and GLE35 addressed by
\cite{Cliver2006}, and his conclusion about these events does not
provide a straightforward key to understanding GLE63. The
time-profiles of the proton fluxes are typical of well-connected
events, with a sharp rise roughly corresponding to the flare peak
followed by a moderately long decay. HEPAD detected a
highest-energy $>700$\,MeV enhancement, which is usual in GLE
events. Each integral channel is dominated by protons in its
lowest-energy part because of their declining spectrum. The time
profiles reflect the energy spectrum of protons, while the path
lengths for different energies might not be identical. The
increasing duration of the lower-energy proton fluxes is mainly
caused by their transport in the interplanetary space, primarily
the velocity dispersion. It is not possible to exclude
contributions from two different accelerators, one of which
operated longer, dominating at lower energies, while another of a
shorter duration dominated at higher energies.

From an analysis of 35 GLEs (without GLE63)
\cite{Miroshnichenko2013} found the presence of a prompt component
with an exponential flux spectrum and a slow power-law component
with an average $\langle{\gamma}\rangle = 4.85 \pm 0.25$. The
authors considered this fact incompatible with exclusive
acceleration of protons by shock waves, expecting a power-law flux
spectrum with $\gamma \approx 2.5$ in this case.

\cite{Mewaldt2012} presented the energy spectra of proton fluences
and other SEP properties during the 16 GLEs of Solar Cycle 23. The
spectra in a range of $\approx 0.1-600$\,MeV are best fit with
double power-laws. On average, the spectra above $\approx 40$\,MeV
in GLE events have a slope of $-3.18$ with $\sigma = 0.83$,
significantly harder than in typical large SEP events ($-4.34$,
$\sigma = 0.77$). The spectral slopes below the break-energies are
similar, being about $-1.25$ on average. The spectrum of the
proton fluence in our event had a slope of $-1.53$ below 32\,MeV
and $-3.14$ at higher energies, close to the average values for
other GLEs. The double power-law spectrum seems to favor the
dominance of shock-acceleration at lower energies and
flare-acceleration at higher energies advocated by
\cite{Miroshnichenko2013}. On the other hand, a double-power-law
can result from a single power-law spectrum affected by
proton-amplified Alfv{\'e}n waves near the Sun (see
\citealp{Mewaldt2012} for a review) or from the acceleration by a
quasi-perpendicular shock \citep{TylkaLee2006}.

To summarize, the properties of the 26 December 2001 proton
enhancement were typical of well-connected GLE events. There are
indications of the contributions from both flare-related and
shock-related sources. However, they can also be interpreted in
terms of a single shock-wave accelerator, being therefore
inconclusive.

\subsection{Some Properties of Heavier Ions}

The SEP events have traditionally been categorized as gradual or
impulsive events (mixed events are also considered). Gradual SEP
events are characterized as long-duration, large, intense events.
They have average ion abundances similar to those of the corona or
solar wind. In contrast, impulsive SEP events are small, have
relatively short durations, can have 1000-fold enhancements in
$^3$He/$^4$He and in heavy elements ($Z>50$)/O relative to the
corona or solar wind, and are associated with solar flares or jets
and Type III radio bursts \citep{Reames2013}. The 26 December 2001
event had a long duration and low $^3$He/$^4$He ratio like gradual
events \citep{Desai2006} and exhibited some properties of
impulsive events.

\cite{Mewaldt2012} examined the Fe/O ratio, which is considered as
a diagnostic of flare material. The Fe/O ratio typical of
impulsive SEPs is about unity, while that of gradual SEPs is
around 0.1. A criterion for Fe-rich GLE events is Fe/O~$\geq
0.268$ \citep{Tylka2005}. The authors of both studies noted that
the Fe-rich GLEs, on average, have much smaller $>30$\,MeV proton
fluences than the Fe-poor GLEs. In our event, the Fe/O ratio in
the range 45 to 80 MeV/nucleon was 0.671, and the $>30$\,MeV
proton fluence was $1.16 \times 10^7$\,protons\,cm$^{-2}$, while
the latter parameter for the 16 GLEs was in the range $(8.02
\times 10^6 - 4.31 \times 10^9)$\,protons\,cm$^{-2}$ with a
logarithmic average of $1.77 \times 10^8$\,protons\,cm$^{-2}$.

At the rise phase of some gradual events, Fe/O is $\approx 1$ and
diminishes afterwards, suggesting a flare-related prompt component
and shock-related slow component. The 26 December 2001 SEP event
also showed this behavior. On the other hand, \cite{Tylka2013}
argued the initial Fe/O enhancement in this event to be a
transport effect, advocating only the shock-related accelerator.
We note, however, that if really so, then the mentioned pattern
established by \cite{Tylka2005} and \cite{Mewaldt2012} between the
event-integrated values of the Fe/O ratio and the $>30$\,MeV
proton fluence did not hold in this event. Thus, invoking the Fe/O
ratio still has not determined the source of SEPs on 26 December
2001.

A promising characteristic of the equilibrium temperature in the
acceleration region is the mean ionic charge state of iron
[$\langle{Q_\mathrm{Fe}}\rangle$]. In 10 out of the 16 GLEs, in
which it was measured, $\langle{Q_\mathrm{Fe}}\rangle$ ranged from
11.7 to 22.1. In our event, $\langle{Q_\mathrm{Fe}}\rangle = 20.7$
corresponds to about 10\,MK, which seems to indicate flare
material. \cite{Mewaldt2012} point out that the highly ionized
$>20$\,MeV/nucleon ions in this event and some others could also
be the result of electron stripping during the acceleration and/or
transport process in a sufficiently dense ambient plasma. The
authors concerned with this effect consider that, in terms of the
traditional concept (\textit{e.g.} \citealp{Reames2009a,
Reames2013}), a CME-driven bow-shock can appear and start
accelerating ions between $\approx 140$ and $\approx 400$\,Mm
above the photosphere. However, recent results listed in
Section~\ref{S-introduction} show that shock-acceleration can
occur in still lower corona, and it looks surprising to us in this
case that electron stripping is not common, so that high
$\langle{Q_\mathrm{Fe}}\rangle \approx 20$ are not always
observed.

Thus, the studies of low to moderate-energy protons and heavier
ions reveal indications of both shock--related and flare--related
contributions in the 26 December 2001 event. However, the latter
are not certain and can be interpreted in different ways. Note
that the results of these studies were interpreted in terms of old
hypotheses, while their update might lead to different
conclusions.

\subsection{Highest-Energy Particles}

\begin{figure} 
  \centerline{\includegraphics[width=0.75\textwidth]
   {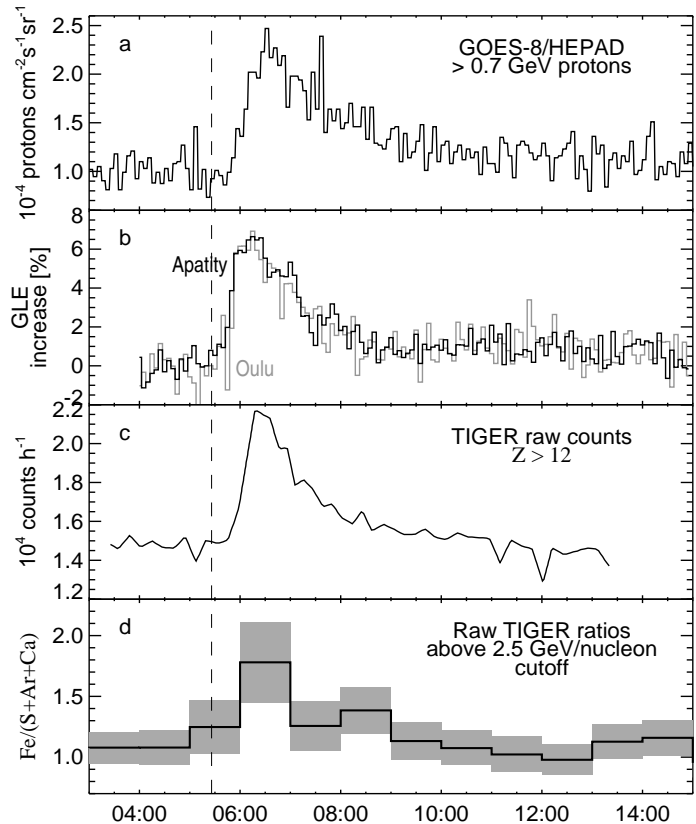}
  }
  \caption{Data on high-energy heavy particles. (a)~GOES-8/HEPAD P11
proton channel $> 700$\,MeV. (b)~GLE recorded with the Apatity
(black) and Oulu (gray) neutron monitors. (c)~Raw record of the
Antarctic TIGER balloon experiment of response to high-energy ions
with $Z \geq 13$. (d)~Raw ratios of heavy ions above the
2.5\,GeV/nucleon cutoff calculated from the TIGER data. The
vertical-dashed line denotes the particle release time
\citep{Reames2009b}.
 }
  \label{F-high-energy}
  \end{figure}

The highest-energy manifestations of SEPs on 26 December 2001 are
shown in Figure~\ref{F-high-energy}. A proxy of ground-level
events is presented by the HEPAD proton channel $> 700$\,MeV in
Figure~\ref{F-high-energy}a. It does not always correspond to a
GLE produced by particles of still higher energies $\gsim 1$\,GeV
\citep{Miroshnichenko2013}. Figure~\ref{F-high-energy}b shows
GLE63 recorded by the Apatity and Oulu neutron monitors. The
vertical-dashed line denotes the solar particle release time
(SPR), 05:20.6~ST\,$\pm 3.7$\,minutes (Solar Time refers to an
event on the Sun, leading UTC by the propagation time of light,
\textit{i.e.} 05:29:00\,UTC\,$\pm 3.7$\,minutes), estimated by
\cite{Reames2009b} from the velocity--dispersion analysis (VDA).

A possible indication of heavy ions accelerated to very high
energies in our event is presented by preliminary data from the
\textit{Trans-Iron Galactic Element Recorder} (TIGER:
\citealp{Geier2003}). TIGER was launched on 21 December 2001 and
flew for about 32 days on a long-duration balloon mission from
McMurdo Base in Antarctica. Being designed mainly to measure the
elemental abundances of galactic cosmic ray nuclei, TIGER observed
the 26 December 2001 GLE event in the $\approx$\,GeV/nucleon
range. Figure~\ref{F-high-energy}c shows raw counts produced by
heavy elements with $Z > 12$ passing through raw triggers of the
device. The shape of the raw TIGER flux is similar to that
measured by neutron monitors in Figure~\ref{F-high-energy}b,
manifesting a common origin of the detected events. No similar
deviations from the data gathered during the other time intervals
of the flight were observed. Figure~\ref{F-high-energy}d shows the
ratios of heavy elements that reached the C0 TIGER Cherenkov
detector. An interesting point here is that the estimated
lower-energy limit of particles that can trigger the C0 detector
is around 2.5\,GeV/nucleon. During the event, the efficiency of
the TIGER track-reconstruction software dropped from $\approx
80\,\%$ to under $\approx 60\,\%$, resulting in an additional dead
time, that might affect the measured ratio \citep{Geier2003}.
Still Figure~\ref{F-high-energy}d shows more iron and presents
unique evidence that heavy ions could be accelerated to such high
energies during this event.

Acceleration of protons and heavier ions in the 26 December 2001
event up to relativistic energies is certain, but their sources
still remain unclear. An additional indication can be found from
statistical relations between the parameters of SEPs, on the one
hand, and those of flares and CMEs, on the other hand.

\subsection{Protons \textit{vs.} 35 GHz Burst and CME Statistics}

A correlation between near-Earth proton enhancements and microwave
bursts has been known for a long time \citep{Croom1971,
CastelliBarron1977, Akinian1978, Melnikov1991}. \cite{Kahler1982}
explained this correlation by the ``big flare syndrome'' (BFS),
\textit{i.e.}, a general correspondence between the energy release
in an eruptive flare and its various manifestations. According to
his idea, SEPs are accelerated by shock waves, while different
parameters of eruptive events should correlate with each other,
independent of any physical connection between them. Supporting
this concept, \cite{Kahler1982} analyzed the correlations between
the peak proton fluxes at 20\,--\,40\,MeV and 40\,--\,80\,MeV in
50 SEP events observed in 1973\,--\,1979, on the one hand, and
microwave data at 8.8\,GHz, 15.4\,GHz, and two lower frequencies,
on the other hand, using the lists of selected parameters. No
proxy of any shock parameters was available. \cite{Kahler1982}
found that the peak proton fluxes correlated with microwaves no
more than with the thermal SXR flare emission. Assuming that
protons are accelerated either by flares or by shocks and not by
both, he favored shock-acceleration. This conclusion caused
skepticism to the correlations between the parameters of SEPs and
microwave bursts.

Studies by \cite{Grechnev2013b, Grechnev2015a} call for rethinking
the role of the BFS. They analyzed the relations between the
parameters of strong microwave bursts $> 1000$\,sfu (1\,sfu =
$10^{-22}$\,W\,m$^{-2}$\,Hz$^{-1}$) at 35\,GHz observed by NoRP in
1991\,--\,2012 and near-Earth proton enhancements $> 100$\,MeV,
both from detailed temporal histories. Gyrosynchrotron emission of
high-energy electrons depends on their parameters, magnetic field
in the source, and its dimensions. These dependencies are
different at frequencies below the turnover frequency of the
gyrosynchrotron spectrum and above it. The turnover frequency also
depends on the parameters listed. Single-frequency data are
therefore ambiguous, being below the turnover frequency in one
event and above it in another. To minimize this ambiguity, a
highest frequency of 35\,GHz was chosen, at which stable regular
observations are available.

In addition to the proton events related to strong microwave
bursts, some big SEPs might have occurred after weaker bursts. A
few additional proton enhancements $> 100$\,MeV with peak fluxes
$J_{100}> 10$\,pfu (1\,pfu =
1\,particle\,cm$^{-2}$\,s$^{-1}$\,sr$^{-1}$) were found, whose
solar source events occurred within the observational daytime in
Nobeyama.

Out of the total set of events, 28 proton enhancements in
1996\,--\,2012 were selected, whose sources were not occulted and
for which data on the corresponding CMEs are listed in the online
CME catalog (\url{cdaw.gsfc.nasa.gov/CME_list/}:
\citealp{Yashiro2004}) based on the observations by the SOHO's
\textit{Large Angle and Spectroscopic Coronagraph} (LASCO:
\citealp{Brueckner1995}). Because the speeds listed in the CME
catalog are measured for the fastest feature, $V_\mathrm{CME}$ for
fast CMEs are most likely related to shock waves
\citep{Ciaravella2006}. The halo shock fronts ahead of fast CMEs
should have the shapes close to spheroidal ones
\citep{Grechnev2011, Grechnev2013a, Grechnev2014b, Kwon2014,
Kwon2015}; thus, the plane-of-the sky speeds measured in the
catalog should not be much different from the modules of their
vectors (``space speeds''), especially in the logarithmic scale.

\begin{figure} 
  \centerline{\includegraphics[width=\textwidth]
   {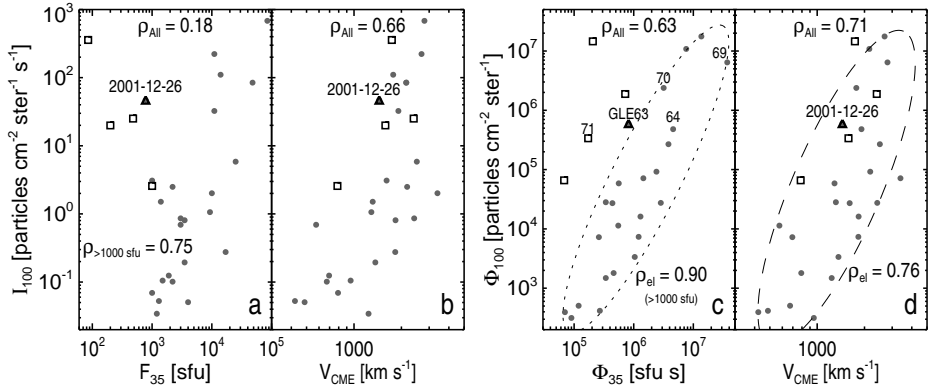}
  }
  \caption{Near-Earth high-energy proton enhancements and strong
microwave bursts recorded by NoRP at 35 GHz in 1996\,--\,2012.
Scatter (log--log) plots present the longitude-corrected
parameters of $>100$\,MeV protons (peak flux $I_{100}$ in the left
pair of panels, total fluence $\Phi_{100}$ in the right pair of
panels) \textit{versus} parameters of microwave bursts at 35\,GHz
(peak flux $F_{35}$ left, fluence $\Phi_{35}$ right) and CME
speed. The Pearson correlation coefficients at the tops of the
panels were calculated for all 28 events ($\rho_\mathrm{All}$),
and those at the bottoms of panels c and d ($\rho_\mathrm{el}$)
are related to the events within the ellipses. The open squares
denote the events with abundant proton outcome. The filled
triangle denotes the 26 December 2001 GLE63 event. The GLE numbers
are indicated at corresponding points in panel c.}
  \label{F-correlations}
  \end{figure}

Figure~\ref{F-correlations} supplements the results of the analysis
by \cite{Grechnev2015a}. Figure~\ref{F-correlations}a shows the
scatter plot of the peak proton flux, $J_{100}$, \textit{vs.} peak
microwave flux, $F_{35}$. The open squares and black triangle (our
event) represent the SEPs with $J_{100}> 10$\,pfu related to the
bursts with $F_{35} \leq 1000$\,sfu. These five points are displaced
from the majority of SEPs denoted by the filled-gray circles, which
show a trend between $F_{35}$ and $J_{100}$ with a Pearson
correlation coefficient of 0.75. This figure is analogous to what
\cite{Kahler1982} presented for lower proton energies and shows a
similar result. For the whole set of events, the correlation
coefficient between $V_\mathrm{CME}$ and $J_{100}$ in
Figure~\ref{F-correlations}b is higher, and the five proton-abundant
events fall mostly within the main cloud of points.

The influence on the proton flux of the processes affecting their
propagation from the source to detector (such as accumulation of
trapped protons and velocity dispersion) can be compensated by
considering their fluence. Figures \ref{F-correlations}c and
\ref{F-correlations}d compare total proton fluences $\Phi_{100}$
with total microwave fluences $\Phi_{35}$ and $V_\mathrm{CME}$.
The broken ellipses enclosing the majority of the events are
plotted by hand. The tilt of the major axis of each ellipse shows
the trend, and its width represents the scatter. Note that the
correlation coefficients rather than the scale-dependent
eccentricities of the ellipses are significant.

Two groups of events show up. The first-group events with $F_{35}
\geq 10^3$\,sfu (gray circles) form a rather narrow cloud  in
Figure~\ref{F-correlations}c within the dotted ellipse with a
correlation coefficient as high as 0.90. The five proton-abundant
events of the second group (with $F_{35} \leq 10^3$\,sfu) remain
isolated, although they approach the main cloud of points. On the
other hand, the main cloud of points within the dashed ellipse in
Figure~\ref{F-correlations}d includes almost all abundant events
(three open squares and the triangle of our event). Their
arrangement nearly along its major axis corresponds to the main
trend. The only exception is a big 8\,--\,9 November 2000 SEP
event (\textit{e.g.} \citealp{Lario2009}). It is located not far
from the ellipse, being much closer to the main cloud of points
than in Figure~\ref{F-correlations}c. The correlation coefficient
for the whole set of events is higher with the CME speed in
Figure~\ref{F-correlations}d than with the microwave fluence in
Figure~\ref{F-correlations}c, supporting a shock-related
contribution.

An apparent interpretation of Figure~\ref{F-correlations}c is that
the well-correlated SEPs of the first group were dominated by the
flare-related acceleration, because the total number of protons
depends on both the intensity and duration of the acceleration
process. The correspondence between these parameters of the
acceleration process and microwave burst is obvious, but not
expected, if protons are accelerated by shock waves far away from a
flare region. The shock-related acceleration seems to dominate in
the proton-abundant events of the second group. Nevertheless, their
location closer to the main cloud of points in
Figure~\ref{F-correlations}c relative to
Figure~\ref{F-correlations}a supports the flare-related contribution
in these events, too.

The asymmetry of flare magnetic configurations causes an
additional scatter in the correlations between microwave bursts
and SEPs. This asymmetry in the 26 December 2001 event reduced the
microwave burst by a factor of two with the same production of
accelerated particles (Article~II).

GLE64, GLE69, and GLE70 fall within the ellipse in
Figure~\ref{F-correlations}c, which indicates their association
with the first group. Detailed studies of the GLE69-related solar
and particle event \citep{Grechnev2008, Klein2014} support the
flare-related source of SEPs. The solar source event of GLE70 was
similar to that of GLE69 \citep{Grechnev2013a} which also supports
the indication of Figure~\ref{F-correlations}c. GLE71 located away
from the ellipse in Figure~\ref{F-correlations}c and within the
ellipse in Figure~\ref{F-correlations}d looks like a
shock-dominated event. GLE63 seems to have significant
contributions from both flare-related and shock-related
accelerations.

The high correlation for the first-group events within the ellipse
in Figure~\ref{F-correlations}c holds over three orders of
magnitude for the microwave fluence and five orders for the proton
fluence, while $V_\mathrm{CME}$ in Figure~\ref{F-correlations}d
range over one order of magnitude. A general pattern expressed by
\cite{Kahler1982} in terms of the BFS appears to be more complex
than the correlation among all parameters in big flares. A general
correspondence between the parameters of flares, CMEs, shock
waves, and SEPs holds over a wide range of their magnitudes.
According to \cite{Dierckxsens2015} and \cite{Trottet2015}, the
shock-related contribution statistically dominates at lower
energies with a major role of flares at higher energies, where
shock-accelerated SEPs also show up \citep{Cliver2006,
Gopalswamy2015}.

Based on these facts and considerations, contributions to the 26
December 2001 SEP event from both flare processes and shock waves
may be expected. We use this assumption as a guideline in our
analysis.

\section{Eruptions}
  \label{S-eruptions}

\begin{figure} 
  \centerline{\includegraphics[width=\textwidth]
   {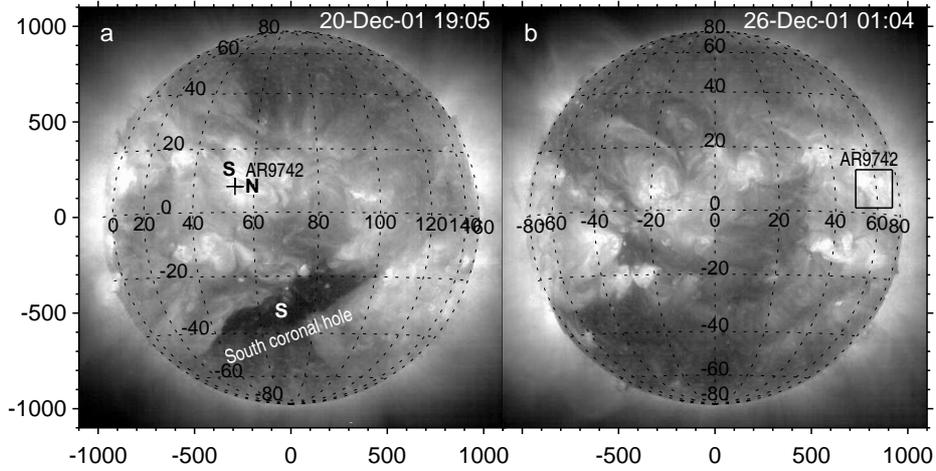}
  }
  \caption{The overall situation on the Sun observed by SOHO/EIT in
284\,\AA. (a)~The Sun on 20 December. The reported position of the
26 December flare in AR\,9742 is denoted by the cross. A large
coronal hole is present in the south hemisphere. Magnetic
polarities S and N in the coronal hole and in AR\,9742 are
indicated. The heliographic grid corresponds to the flare
occurrence time. (b)~The Sun on 26 December before the event. The
black frame outlining the flare region corresponds to the field of
view in Figure~\ref{F-flare_regions}. The axes indicate the
distance from solar disk center in arcseconds.}
  \label{F-eit284}
  \end{figure}

The eruptive flare on 26 December occurred in AR\,9742 not far
from the west limb (N08\,W54). Figure~\ref{F-eit284}a shows an EIT
284\,\AA\ image observed on 20 December, 130 hours before the
event. The cross denotes the reported position of the flare.
AR\,9742 had a $\beta \gamma$ magnetic configuration. Approximate
positions of the ends of the erupted flux rope and their magnetic
polarities revealed by the flare ribbons and a magnetogram
produced by \textit{Michelson Doppler Imager} (MDI:
\citealp{Scherrer1995}) on SOHO (see Article~II) are denoted S and
N. A large S-polarity coronal hole opposite to AR resided in the
southern hemisphere.

\begin{figure} 
   \centerline{\includegraphics[width=\textwidth]
    {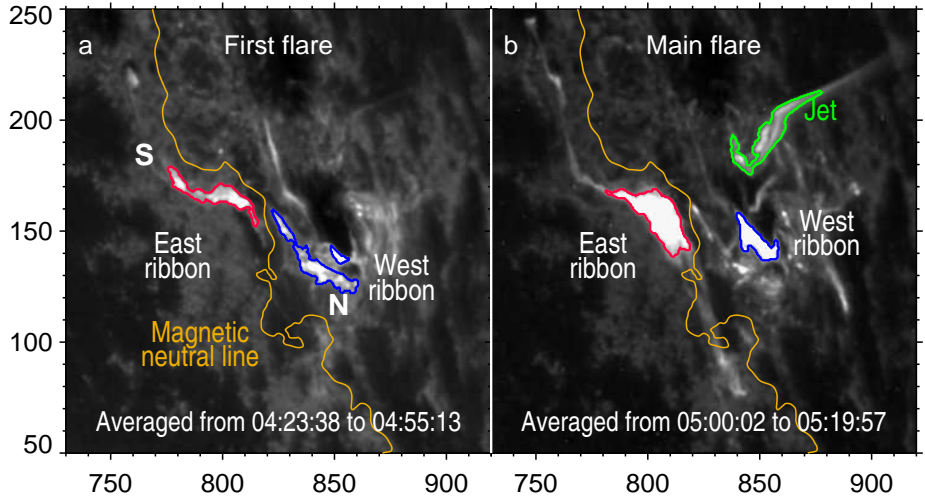}
    }
   \caption{Two parts of the two-ribbon flare represented by
TRACE 1600\,\AA\ images averaged during the first flare (a) and
main flare (b). The blue contours outline the west flare ribbons
in the sunspot. The red contour outlines the east ribbon in a
weaker-field region. The orange contour traces the magnetic
neutral line computed from the SOHO/MDI magnetogram observed on 26
December at 04:51 (see Article~II). Magnetic polarities S and N
are indicated. The green contour in panel b outlines the brightest
portion of the jet and a part of its base. The axes indicate the
distance from solar disk center in arcseconds.}
   \label{F-flare_regions}
\end{figure}

The EIT 284\,\AA\ image in Figure~\ref{F-eit284}b presents the Sun
on 26 December a few hours before the event. The southern coronal
hole hidden by bright coronal structures rotated to the limb (it
was visible again on 15\,--\,19 January 2002, when its area
decreased). The black frame corresponds to the field of view in
Figure~\ref{F-flare_regions}, which shows the flare observed by
TRACE in 1600\,\AA\ (see Article~II for more details). The whole
event consisted of the first flare (Figure~\ref{F-flare_regions}a)
and the main flare (Figure~\ref{F-flare_regions}b).

\begin{figure} 
   \centerline{\includegraphics[width=0.75\textwidth]
    {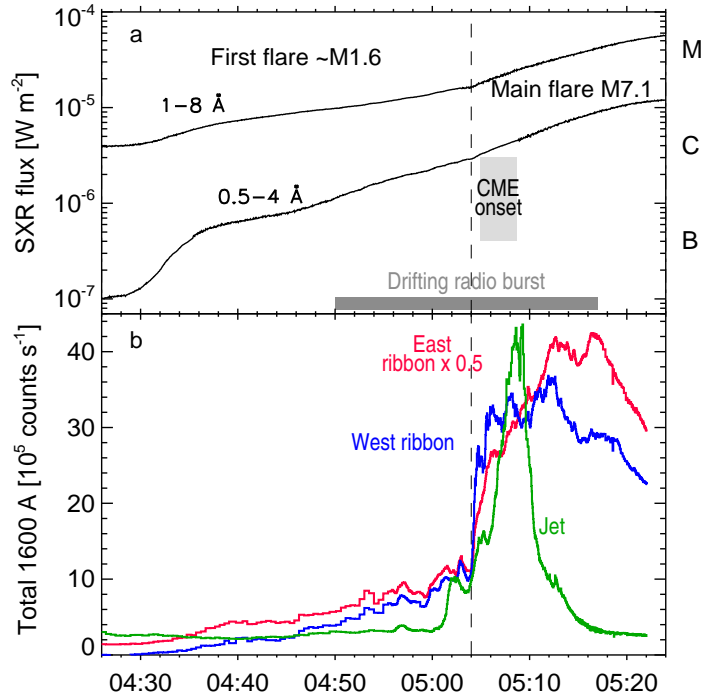}
   }
   \caption{Flare light curves recorded by GOES in soft X-rays (a) and
those computed from TRACE 1600\,\AA\ images in
Figure~\ref{F-flare_regions} over the flaring regions presented
with the corresponding colors (b). Two flare parts are separated
by the dashed-vertical line. The gray bars in panel a represent
the observation interval of a slowly drifting radio burst and the
CME onset time extrapolated to the position of AR\,9742.}
   \label{F-flare_light_curves}
\end{figure}

\subsection{First Eruption}

The first, most likely eruptive flare started from the appearance
of two long, thin, strongly sheared ribbons
(Figure~\ref{F-flare_regions}a). The west ribbon was close to the
sunspot. The east ribbon was located in moderate magnetic fields.
The SXR flux in Figure~\ref{F-flare_light_curves}a started to rise
after 04:30 and reached a GOES importance of about M1.6 at 05:04.
Figure~\ref{F-flare_light_curves}b shows the time-profiles
computed from the TRACE 1600\,\AA\ images over the major regions
outlined in Figure~\ref{F-flare_regions} with corresponding
colors. Both ribbons gradually brightened by 05:04.

\begin{figure} 
   \centerline{\includegraphics[width=\textwidth]
    {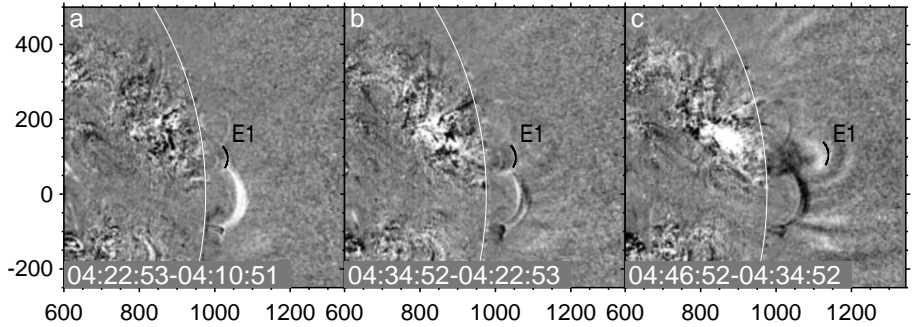}
   }
   \caption{First eruption in running-difference EIT 195\,\AA\ images.
(a,~b)~Slow expansion of coronal structures in the initiation
phase. The black arc denoted E1 outlines the visible top of the
rising structures. (c)~Manifestations of the eruption in the
stretched coronal structures. The axes indicate the distance from
solar disk center in arcseconds.}
   \label{F-first_eruption}
   \end{figure}

A slowly drifting Type II and/or Type IV burst in an interval
marked in Figure~\ref{F-flare_light_curves}a could only be caused
by an expanding ejecta or wave from AR\,9742, which started, at
least, ten minutes before the main fast CME. Manifestations of the
first eruption in running-difference EIT 195\,\AA\ images are
presented in Figure~\ref{F-first_eruption}. The top of a faint
off-limb loop-like feature E1 in the pre-eruption image
(Figure~\ref{F-first_eruption}a) is outlined by the black arc.
This top is displaced slightly in the initiation phase at 04:34:52
in Figure~\ref{F-first_eruption}b and strongly in
Figure~\ref{F-first_eruption}c at 04:46:52, when it accelerated
and brightened. Its lift-off apparently stretched closed coronal
structures lying above. A dark dimming-like region above the limb
started developing behind E1. The images in
Figure~\ref{F-first_eruption} provide the projected heights of E1
at the three times.

\subsection{Main Eruption}
 \label{S-major_eruption}

The main flare started at 05:04, close to the estimated CME onset
time (light-gray bar in Figure~\ref{F-flare_light_curves}a; see
Article~II). Emissions from both ribbons in 1600\,\AA\ and
microwaves strongly increased. The west ribbon reached the sunspot
umbra and partly covered it. The east ribbon lengthened and
broadened into weaker-field regions. The SXR flux strengthened and
reached an M7.1 importance at 05:40.

\begin{figure} 
   \centerline{\includegraphics[width=0.75\textwidth]
    {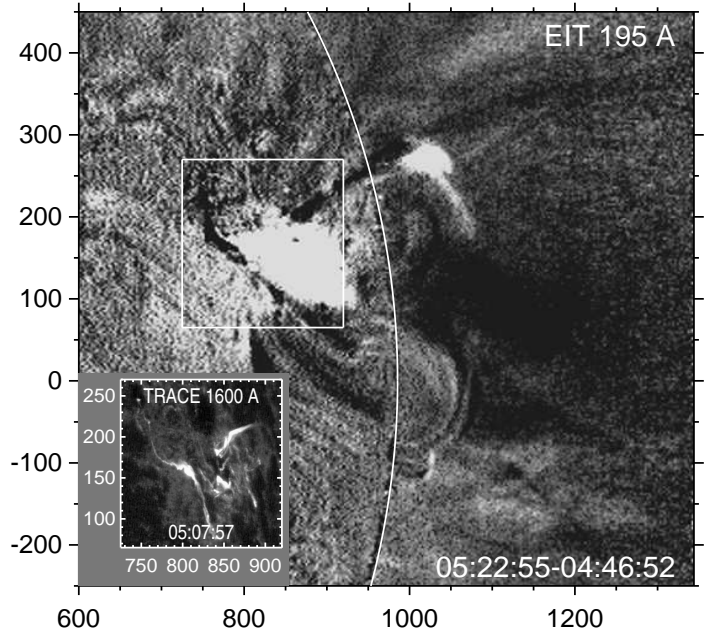}
   }
   \caption{Traces of the main eruption in the EIT 195\,\AA\
difference image. The inset presents a TRACE 1600\,\AA\ image near
the flare peak. Its actual position is denoted by the white frame.
The axes indicate the distance from solar disk center in
arcseconds.}
   \label{F-major_eruption}
   \end{figure}

The EIT 195\,\AA\ difference image in
Figure~\ref{F-major_eruption} reveals the traces of the associated
main eruption, which occurred between 04:47 and 05:22. The onset
time of the main flare and CME falls within this interval. Coronal
structures are strongly disturbed. A large dimming surrounded by
stretched loops appears above the limb. The flare configuration is
not recognizable in the EIT image because of the low brightness
threshold applied to detect faint surrounding features. A
two-ribbon structure and bright jet are visible in a
high-resolution TRACE 1600\,\AA\ image in the inset, whose actual
position is denoted by the white frame.

\subsection{Jet}
 \label{S-jet}

\begin{figure} 
   \centerline{\includegraphics[width=\textwidth]
    {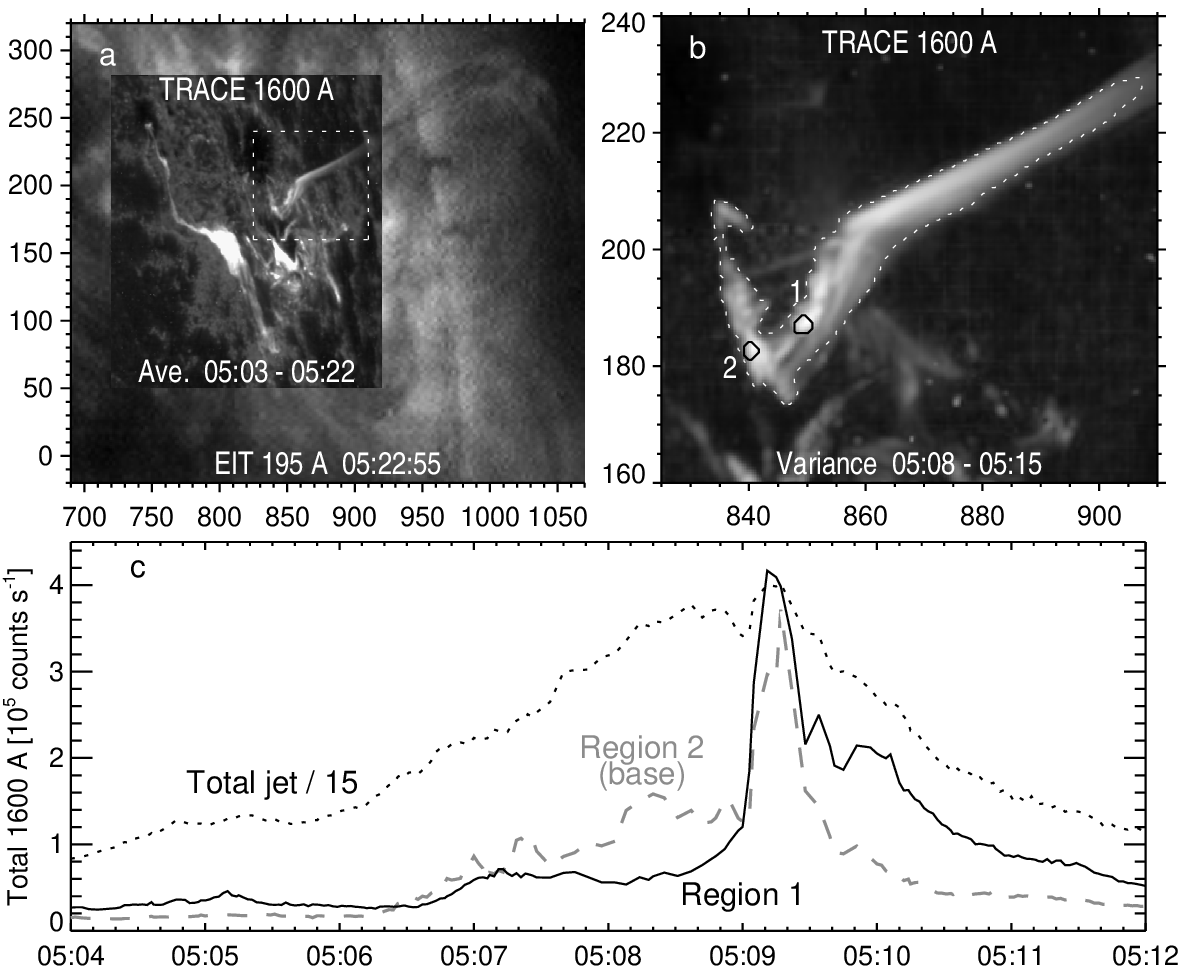}
   }
   \caption{(a)~The overall configuration of the jet in an
EIT 195\,\AA\ image with an inserted averaged TRACE 1600\,\AA\
image. A jet visible in 1600\,\AA\ occurred in a long loop-like
structure visible in 195\,\AA. (b)~Variance map of the jet
produced from the TRACE 1600\,\AA\ images during 05:08\,--\,05:15.
The field of view corresponds to the white dotted frame in panel
a. Labels 1 and 2 denote sharply brightened regions. The axes in
panels a and b indicate the distance from solar disk center in
arcseconds. (c)~Temporal profiles over the whole jet within the
dotted contour (reduced by a factor of 15) and those of the small
regions 1 and 2 denoted in panel b.}
   \label{F-eit_trace}
   \end{figure}

A jet (green in Figures \ref{F-flare_regions} and
\ref{F-flare_light_curves}b) appeared after 05:06 from a
funnel-like structure, while brightenings ran along its circular
base. The time-profile of the jet in 1600\,\AA\ reached a peak at
about 05:09, being as short as three minutes at half-height.
Figure~\ref{F-eit_trace} presents the jet-like eruption. A
combination of an EIT 195\,\AA\ image and an averaged TRACE
1600\,\AA\ image in Figure~\ref{F-eit_trace}a reveals a
large-scale configuration, where the jet occurred.
Figure~\ref{F-eit_trace}b presents a small-scale configuration,
from which the jet emanated. This is an enlarged variance image of
the jet computed from the TRACE 1600\,\AA\ images in an interval
from 05:08 to 05:15. This image represents all changes occurring
in this interval according to their statistical contributions
\citep{Grechnev2003vm}.

The coronal configuration in Figure~\ref{F-eit_trace}b resembles
an inverted funnel. Such funnels appear above photospheric
magnetic islands inside opposite-polarity regions and contain
coronal null points \citep{Masson2009, Meshalkina2009}. The
presence of a magnetic island at the photospheric base of the jet
was revealed in Article~II. A long tube-like extension (dark in
the EIT image in Figure~\ref{F-eit_trace}a) connects the ring base
of the funnel with a remote magnetically conjugate region far away
from AR\,9742. The magnetic structure of a small flux rope
erupting inside a funnel cannot survive when passing a null point
\citep{Uralov2014}, and released plasma flows out as a jet
\citep{Filippov2009}. Jet-like eruptions in such configurations
are characterized by ring-like ribbons, brightenings running along
them, and impulsive temporal profiles.

Two compact brightest regions 1 and 2 in Figure~\ref{F-eit_trace}b
exhibited the largest variations. The temporal profiles computed
over the whole jet and regions 1 and 2 are shown in
Figure~\ref{F-eit_trace}c. They demonstrate that coronal region~1
and region~2 in the base of the jet exhibited a simultaneous
brightening as short as 20\,seconds, suggesting a sharp impulsive
energy release at about 05:09. This pulse, the preceding collision
of the flux rope with the separatrix surface of the funnel, and a
pressure pulse produced in the bend of the long tube-like
structure by the injected dense material of the jet result in a
strong impulsive disturbance excited by the jet around 05:09. The
shock-wave excitation by a similar jet-like eruption was
demonstrated previously by \cite{Meshalkina2009} and
\cite{Grechnev2011}.

In summary, the whole event comprised three eruptions. The first
eruption indicated by the first two-ribbon flare and a
slowly-drifting radio burst occurred around 04:35. The second, main
eruption associated with the fast CME occurred around 05:04. The
third, jet-like eruption was actually observed around 05:09.

\section{Drifting Radio Bursts}
 \label{S-radio bursts}

To find further information about the eruptions and shock wave in
our event, we consider dynamic radio spectra in a wide frequency
range. The structures visible in dynamic spectra reveal
non-thermal electrons streaming along open magnetic fields (Type
III bursts), tracing the fronts of shock waves (Type II bursts),
or confined in quasi-static or expanding magnetic structures (Type
IV bursts).

\subsection{Type II Bursts}
 \label{S-type_II}

To analyze Type II bursts, we use our technique to outline their
trajectory verified in studies of several events. The trajectory
is governed by the plasma density distribution in the way of a
propagating shock wave. A freely propagating blast-wave-like
shock, which spends energy to sweep up the plasma with a radial
power-law density falloff, $n(x) \propto x^{-\delta}$ ($x$ is the
distance from the eruption center), has a power-law kinematic
behavior, $x(t) \propto t^{2/(5-\delta)}$ \textit{vs.} time [$t$]
\citep{Grechnev2008shocks}. The power-law density model, $n(h) =
n_0(h/h_0)^{-\delta}$, with [$h$] being the height above the
photosphere, $n_0 = 4.1 \times 10^8$\,cm$^{-3}$, $h_0 = 100$\,Mm,
and $\delta = 2.6$ is close to the equatorial Saito model
\citep{Saito1970} at $h \geq 260$\,Mm, providing a steeper density
falloff at lower heights. The low-corona density increase
corresponds to strongly disturbed conditions just before the
appearance of the wave.

The expected trajectory of a Type II burst caused by the passage
of the shock front through a structure with a decreasing density
is a gradual monotonic curve. It has a steep onset and decreasing
frequency drift with a convexity governed by the
$\delta$-parameter. The wave onset time [$t_0$] usually
corresponds to the rise of the HXR or microwave burst or precedes
it by up to two minutes.

In practice, we choose a reference point on the dynamic spectrum
at time $t_1$, calculate a corresponding distance $x_1$ from our
density model, and we adjust $t_0$ and $\delta$ (typically $\delta
\approx 2.5 - 2.9$) in sequential attempts to reach best fit to
bright Type~II signatures of a trajectory calculated from the
equation $x(t) = x_1[(t-t_0)/(t-t_1)]^{2/(5-\delta)}$ and plotted
on top of the dynamic spectrum. Their visual comparison provides a
typical accuracy of about 0.01 for $\delta$ and within one minute
for $t_0$. We also use this approximation to fit different wave
signatures such as ``EUV waves'' and leading edges of fast CMEs
\citep{Grechnev2011, Grechnev2013a, Grechnev2014b, Grechnev2015b,
Grechnev2016}. Note that the dynamic spectrum alone does not allow
us to refer the density to a certain height, being basically
insensitive to a constant multiplier [$n_0$] of any density model.

  \begin{figure} 
  \centerline{\includegraphics[width=\textwidth]
   {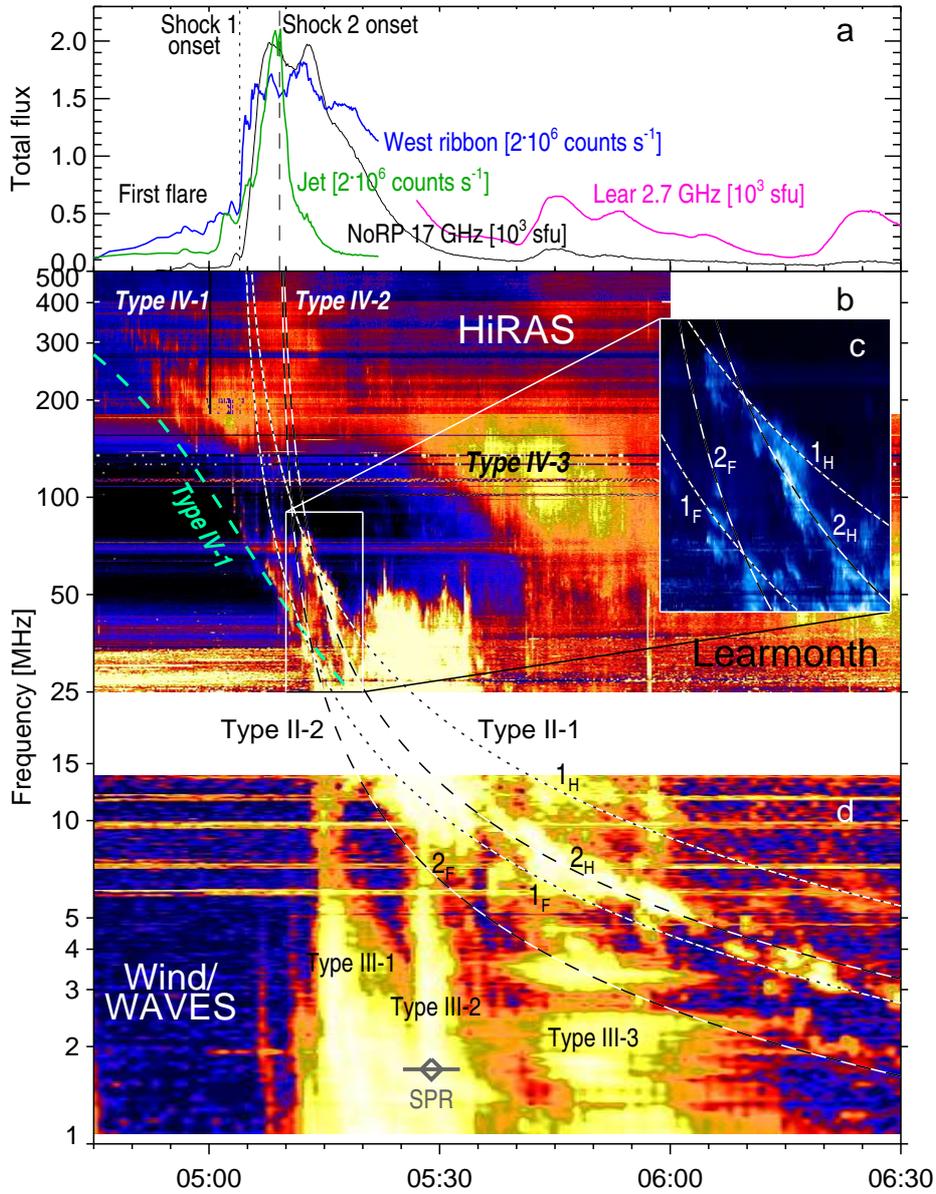}
  }
  \caption{Combined dynamic spectrum (b and d) in comparison
with the flare light curves (a). The vertical-broken lines in
panel a represent the onset times of shock 1 ($t_{01} =
$~05:04:00) and shock 2 ($t_{02} = $~05:09:10). Inset (c) presents
an enlarged part of the Learmonth spectrogram denoted by the white
frame. The slowly drifting radio bursts are outlined with
calculated trajectories. The dashed-green curve (not shown in the
inset) outlines the low-frequency envelope of Type IV-1 above
25\,MHz. Traces of Type II-1 are outlined by the pair of dotted
lines ($\delta_1 = 2.65$), and those of Type II-2 are outlined by
the paired dashed lines ($\delta_2 = 2.54$). The gray diamond at
the bottom labeled SPR denotes the estimated SPR time with an
uncertainty represented by the gray bar \citep{Reames2009b}.}
  \label{F-dynamic_spectrum}
  \end{figure}

Figure~\ref{F-dynamic_spectrum} presents a dynamic spectrum
consisting of a \textit{Hiraiso Radio Spectrograph} (HiRAS)
spectrogram above 180\,MHz and Learmonth data below 180\,MHz in
Figure~\ref{F-dynamic_spectrum}b, and a spectrogram produced by
the Rad2 receiver of the \textit{Radio and Plasma Wave
Investigation} (WAVES: \citealp{Bougeret1995}) on the
\textit{Wind} spacecraft in Figure~\ref{F-dynamic_spectrum}d.
Figure~\ref{F-dynamic_spectrum}a also shows a microwave burst at
17\,GHz (NoRP: black) and 2.7\,GHz (Learmonth: pink, late part
only) as well as the total emission in 1600\,\AA\ from the west
ribbon and jet (same as in Figure~\ref{F-flare_light_curves}b).

An enlarged part of the Learmonth spectrogram in the inset
(Figure~\ref{F-dynamic_spectrum}c; the white frame denotes its
actual position) reveals two harmonic pairs of Type~II lanes
crossing each other. The pair ($1_\mathrm{F}$, $1_\mathrm{H}$),
outlined by the dotted lines, is band-split. The pair
($2_\mathrm{F}$, $2_\mathrm{H}$), outlined by the dashed lines,
has a faster frequency drift. Adjustment of the wave onset time
and density falloff exponent for each of the two paired bands
indicates their relation to two different shock waves following
each other. The first shock wave started at $t_{01} = $~05:04:00
($\delta_1 = 2.65$) and was caused by the main eruption, while the
second shock wave ($t_{02} = $~05:09:10, $\delta_2 = 2.54$) was
produced by the jet (\textit{cf.}
Figure~\ref{F-dynamic_spectrum}a).

  \begin{figure} 
  \centerline{\includegraphics[width=\textwidth]
   {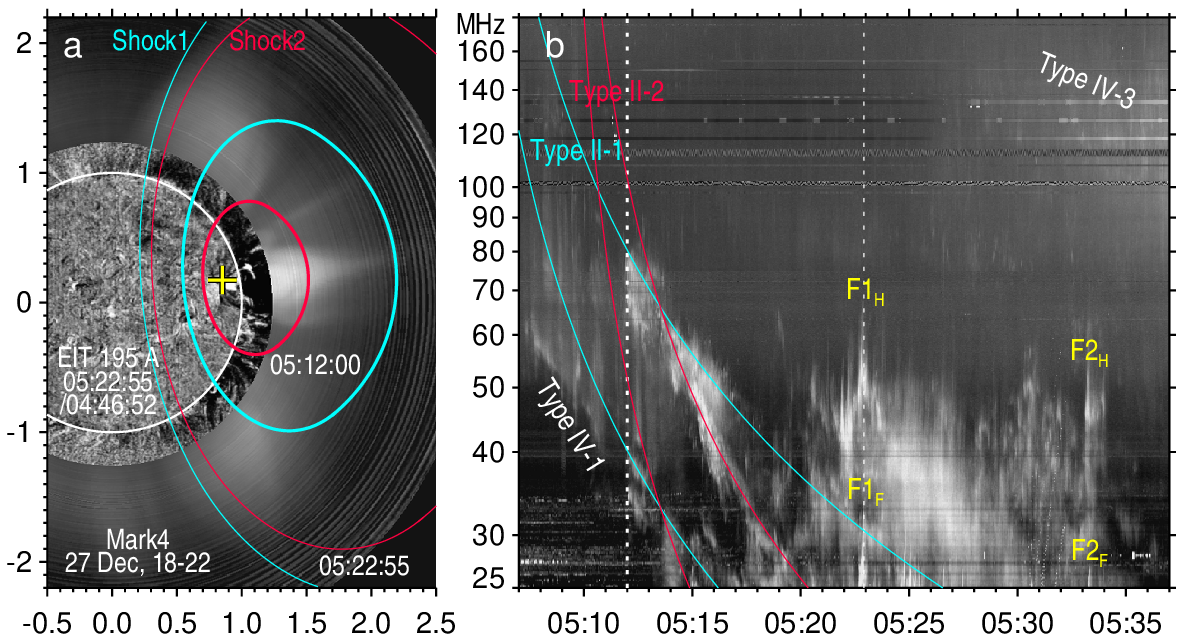}
  }
  \caption{Two shock fronts in the corona (a)
and the structure of the Type~II burst (b). (a)~A collage composed
from an EIT 195\,\AA\ image ratio and a white-light image from the
Mark\,4 coronameter (MLSO). The color curves represent the global
fronts of shock~1 (blue) and shock~2 (red) at 05:12:00 (thick) and
05:22:55 (thin). The yellow cross marks the eruption center. The
white circle corresponds to the solar limb. The axes indicate
solar radii from solar disk center. (b)~The dynamic spectrum
(Learmonth) with the main trajectories of the Type~II-1 (blue) and
Type~II-2 (red) bands. F1$_\mathrm{H}$, F1$_\mathrm{F}$ and
F2$_\mathrm{H}$, F2$_\mathrm{F}$ are harmonically related
features. The vertical-dotted lines denote the times of the shock
fronts in panel a.}
  \label{F-mark4_spectrum}
  \end{figure}

Figure~\ref{F-mark4_spectrum}a shows a schematic of the two shock
fronts at two times. The background is an EIT 195\,\AA\ image
ratio between 05:22:55 and 04:46:52 inserted into an averaged
white-light image from the Mark\,4 coronameter (MLSO) observed on
27 December from 18 to 22\,UT. The thick fronts correspond to the
onset of Type~II-1 (05:12:00) and the thin fronts correspond to
the observation time of the EIT 195\,\AA\ image (05:22:55).
Because of the lack of data required for correct reconstruction of
the wave fronts, their shape is back-extrapolated from LASCO
images (Section~\ref{S-CME}) without crossing the solar surface,
and their size is plotted roughly.

The Learmonth spectrogram in Figure~\ref{F-mark4_spectrum}b shows
details of the Type~II burst, which is saturated in
Figure~\ref{F-dynamic_spectrum}. Its bands partly shift after
05:18 to higher frequencies like an inverse N, indicating that a
part of the shock front entered a denser region (see
\citealp{Grechnev2011, Grechnev2014b}). The Mark\,4 image in
Figure~\ref{F-mark4_spectrum}a really shows on the north and south
broad dense regions, whose continuations should cross the solar
disk. Two narrow-band harmonically related pairs F1$_\mathrm{H}$,
F1$_\mathrm{F}$ and F2$_\mathrm{H}$, F2$_\mathrm{F}$ most likely
belong to the Type~II burst overlapping with Type~IV and different
emissions from 05:20 to 05:34. A Type~III-like emission visible at
25\,--\,60\,MHz between 05:24 and 05:28 has a different origin,
because its enlarged structure shows irregular drifts. A later
structure does not resemble Type~III emission.

The EIT image ratio in Figure~\ref{F-mark4_spectrum}a reveals a
large off-limb dimming visible behind the wave front. The dimming
represents density depletion caused by plasma outflow after the
passage of the shock wave, which upset the vertical hydrostatic
equilibrium of the stratified atmosphere. The plasma flow behind
the shock front swept away and stretched magnetic structures,
increasing their volume. The plasma density near the solar surface
dropped, while its height profile stretched in the streamer above
AR\,9742 visible in the Mark\,4 image as well as a large dimmed
corona. Thus, the second shock wave propagated in a modified
corona with a flatter density profile, $\delta_2 < \delta_1$.

The drift rate [$\mathrm{d}f/\mathrm{d}t$] of the fundamental
emission can be found from an equation $f = f_\mathrm{P}(h, t)$
with $f_\mathrm{P}$ being the plasma frequency in the corona at a
height $h$ and time $t$ corresponding to a portion of the shock
front responsible for a Type~II burst. Differentiation of this
equation gives $\mathrm{d}f/\mathrm{d}t = \partial f_\mathrm{P} /
\partial t + V_\mathrm{shock}\, \partial f_\mathrm{P} / \partial
h\, \cos^m \alpha,$ where $V_\mathrm{shock}$ is a phase speed of
the shock front and $\alpha$ the angle between $\nabla
f_\mathrm{P}$ and the shock normal. Here $m = +1$, if the same
portion of the shock front produces the Type~II burst all of the
time, and $m = -1$, if it is emitted by the intersection of the
shock front with a coronal ray. If the coronal distribution
$f_\mathrm{P} = f_\mathrm{P}(h)$  is stationary and $\alpha = 0$,
then, with $f_\mathrm{P} \approx 9 \times 10^3\, n^{1/2}$ and $n =
n_0(h/h_0)^{-\delta}$, we get $f' = \mathrm{d}f/\mathrm{d}t
\approx V_\mathrm{shock}\,
\partial f_\mathrm{P} / \partial h = -\delta V_\mathrm{shock}\, f
(f/f_\mathrm{P0})^{2/\delta} / (2h_0)$. With $f_\mathrm{P0} =
f_\mathrm{P}(n_0) = 182$\,MHz and the frequency-drift rates of the
fundamental emission for shock~1 [$f'_1$] and shock~2 [$f'_2$] at
their intersection in Figures \ref{F-dynamic_spectrum} and
\ref{F-mark4_spectrum}b (05:13:30, $f \approx 33$\,MHz), we estimate
an instantaneous ratio of their speeds at that time
$V_\mathrm{shock\,2}/V_\mathrm{shock\,1} \approx
(\delta_1/\delta_2)(f'_2/f'_1) \times (f/f_\mathrm{P0})^{(2/\delta_1
- 2/\delta_2)} \approx (2.65/2.54)\times 2 \times (33/182)^{-0.033}
\approx 2$. The height of the Type~II-2 source was different from
that of Type~II-1 at that time, because shock~2 propagated in the
corona modified by shock~1.

Continuations of the trajectories in
Figure~\ref{F-dynamic_spectrum} to the frequencies $< 25$\,MHz are
not obvious because of the gap between the HiRAS and Learmonth and
\textit{Wind}/WAVES bands and complex structures of the Type II
bursts. The identification of the bands in the \textit{Wind}/WAVES
spectrogram is not guaranteed, while the calculated trajectories
match the actual evolution of the frequency drift. The
\textit{Wind}/WAVES spectrogram shows, at least, three Type II
bands, confirming the presence of two shocks. Two shock waves
following each other within a few minutes were observed previously
\citep{Grechnev2011, Grechnev2013a}.

The Type II emission in this event was observed up to a very low
frequency of about 150\,kHz. It is a long-standing issue if Type
II emissions observed in the decametric/hectometric (DH) range and
at still longer waves, also termed interplanetary (IP) Type II
events, can be extensions of metric Type II bursts. An apparent
mismatch between the trajectories of the former and latter events
has been considered as an indication of different origins of
responsible shocks.

In this view, \cite{CaneErickson2005} examined several events and
estimated that ``$<15\,\%$ of the Type IIs extending below 15\,MHz
actually extend below 5\,MHz and that the lowest frequency extent
is about 1\,MHz''. The authors found ``no clear example of a
metric Type II burst that extends continuously down in frequency
to become an IP Type II event'' at lower frequencies. The 26
December 2001 event was specially considered by the authors with
the conclusion ``like other events, there is a disjoint in
frequency between the Type II burst and emission likely to be
related to the CME shock [\textit{i.e.} IP Type II event]''.

 \begin{figure} 
  \centerline{\includegraphics[width=\textwidth]
   {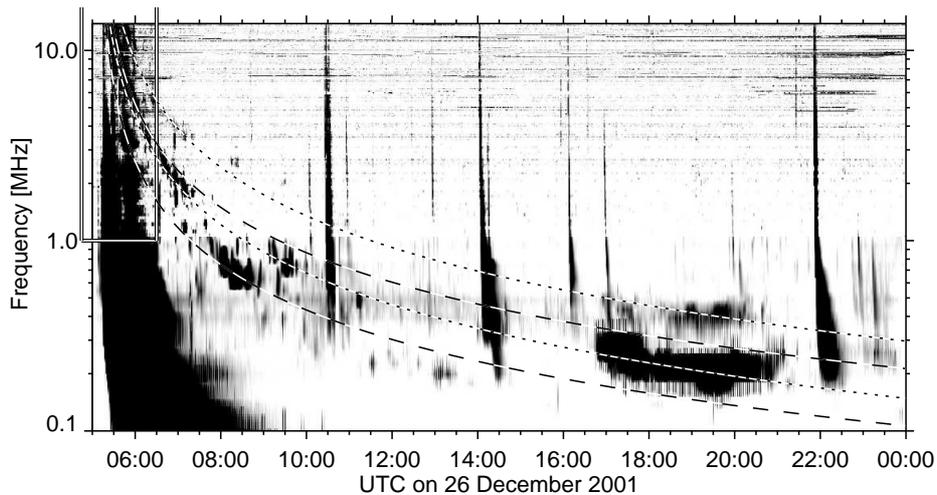}
  }
  \caption{Interplanetary Type II event in a \textit{Wind}/WAVES
Rad2 + Rad1 spectrogram (adapted from \citealp{CaneErickson2005}.
Courtesy H.V.~Cane). The pairs of broken lines represent the same
calculated trajectories of two shock waves following each other,
shock~1 (dotted) and shock~2 (dashed), as in
Figure~\ref{F-dynamic_spectrum}. The white-on-black frame represents
the range shown in Figure~\ref{F-dynamic_spectrum}.}
  \label{F-type_II_Cane}
  \end{figure}

To understand if our analysis can shed light on this problem, we
extend our outline of the two shock waves from
Figure~\ref{F-dynamic_spectrum} up to 100\,kHz. Because
\cite{CaneErickson2005} were concerned about the proper frequency
scaling of concatenated spectrograms (here \textit{Wind}/WAVES
Rad2 + Rad1), we have taken their Figure~13 and plotted over it
our trajectories found from the metric Type II bursts without any
additional adjustment. Figure~\ref{F-type_II_Cane} shows the
result. The white-on-black frame in the top-left corner represents
the field of view in Figure~\ref{F-dynamic_spectrum}. Continuation
of the metric Type~II burst is visible until 10:00. Then the
Type~II emission reappeared at about 17:00 along the same
calculated trajectories. This correspondence confirms its
persistent origin. It is not clear if the Type~II emission was
interrupted because of unfavorable conditions for its generation
or propagation issues.

Although the Type II bands are not continuous and their
identification is not guaranteed, the calculated trajectories
correctly reproduce the actual evolution of the frequency drift
throughout the event in the whole frequency range without any
frequency mismatch between the metric and IP Type II emission. The
impression of a mismatch was probably caused by a complex
structure of the radio emission with gaps between the Type II
portions observed and the presence of misleading features, which
might be irrelevant to the main trajectories.

Figure~\ref{F-type_II_Cane} also leads to the following
conclusions: i)~The Type~II emission in the whole range where it
was observed, from about 80\,MHz to about 150\,kHz, was due to the
same shock wave, which was excited by the eruption in AR\,9742
during the flare; ii)~The Type II bands and blobs in the
0.5\,--\,5\,MHz range between 06:30 and 10:00 corresponding to
different dotted and dashed trajectories certify the presence of
two different sources of the Type~II emission. This fact rules out
a popular idea relating the Type~II source to the bow-shock ahead
of the CME nose. At least one of the radio sources must be located
at a flank of the shock wave, because two bow-shocks cannot be
driven by a single piston. Most likely, two lateral
blast-wave-like shocks coexisted, at least until 09:00, while
ahead of the CME they merged into a single stronger shock
\citep{Grechnev2011}.

\subsection{Type IV Bursts}
 \label{S-type_IV}

\begin{figure} 
  \centerline{\includegraphics[width=\textwidth]
   {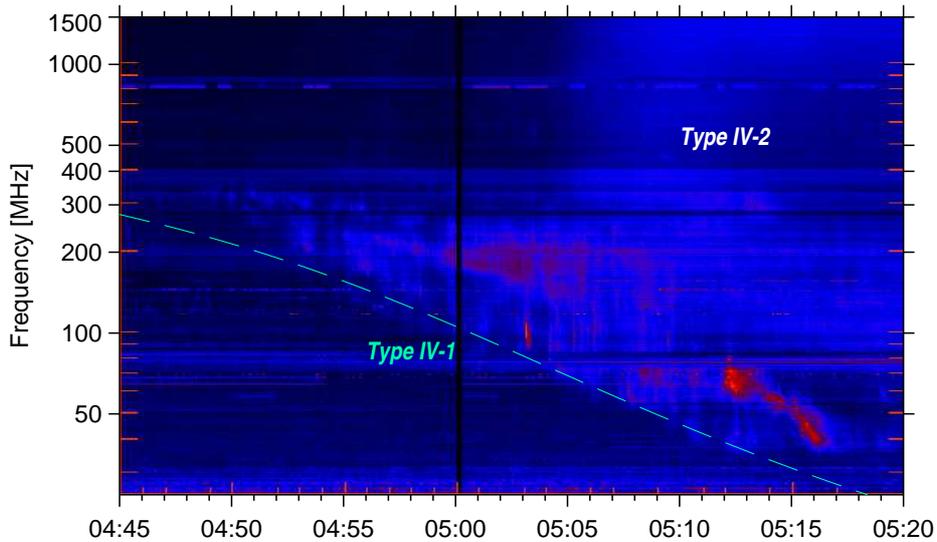}
  }
  \caption{Type IV bursts in the HiRAS spectrogram. The dashed-green
curve (same as in Figure~\ref{F-dynamic_spectrum}b) outlines the
low-frequency envelope of Type~IV-1 according to the height--time
plot in Figure~\ref{F-rope_plots}a. }
  \label{F-type_IV}
  \end{figure}

A slowly drifting burst visible in
Figure~\ref{F-dynamic_spectrum}b from 04:50 until 05:13 outlined
from below by the dashed-green line resembles a Type II burst. So
it was reported by some observatories and considered by
\cite{Nitta2012}. However, the convexity of its trajectory with an
increasing drift rate is opposite to those of Type~II-1 and
Type~II-2. The structure of this burst is better visible in
Figure~\ref{F-type_IV}, which presents an adapted HiRAS
spectrogram (\textsf{2001122605.gif}) accessible at
\url{sunbase.nict.go.jp/solar/denpa/hirasDB/Events/2001/}.

At first glance, the burst comprised a pair of bands with a
frequency ratio of 1.8 (the bright red feature visible from 05:12
to 05:16 between 40 and 80\,MHz is due to Type~IIs). Its spectrum
is cut off towards lower frequencies, as expected for a population
of confined electrons. The high-frequency cutoff has a less
pronounced drift, if any, so that the bandwidth of this burst
increases, which does not resemble a Type~II burst. This burst,
Type~IV-1, suggests emission from an electron population confined
in an expanding magnetic structure.

The second broadband burst in Figure~\ref{F-type_IV}, Type~IV-2,
started about 05:05 in association with the main eruption. Its
quasi-stationary high-frequency part, probably related to the
flare arcade, extended up to $> 2000$\,MHz and lasted until 05:27.
The HiRAS spectrogram suggests a drift of its low-frequency part
to lower frequencies (invisible in
Figure~\ref{F-dynamic_spectrum}b). A slowly drifting burst
Type~IV-3 superposed on a Type~III group appears in
Figure~\ref{F-dynamic_spectrum}b much later, probably due to
emission from the structures well behind the CME leading edge.

Analysis of Type~IV-2 and Type~IV-3 is hampered by their poor
visibility and overlap with different structures. We focus on
Type~IV-1. This slowly drifting burst evidences a moving source. The
moving radio source is often observed along the extrapolated
trajectory of an erupting prominence (see, \textit{e.g.},
\citealp{McLean1973, Klein2002}). By relating a drifting Type~IV
burst to the observed expansion of an SXR source,
\cite{Grechnev2014b} reconstructed its kinematics in a time interval
exceeding imaging observations.

It is not possible to use here either the density model or
kinematics estimated from the analysis of Type~II bursts, unlike
the bow-shock regime, when the kinematics of the piston and shock
is similar. The low-cutoff frequency of a Type~IV burst is
controlled by plasma in an expanding dense structure, while the
frequency of a Type~II burst is governed by its environment. The
structure responsible for a Type~IV burst accelerates starting
from a small velocity; a shock wave responsible for a Type~II
burst starts from the fast-mode speed ($>10^3$\,km\,s$^{-1}$ above
an active region) and then decelerates. Most likely, the first
eruption responsible for Type~IV-1 has not produced a shock wave.

Following the approach of \cite{Grechnev2014b}, we assume that the
frequency drift reflects the decreasing density in an expanding
volume with a size $r$ and relate the low-cutoff frequency to the
plasma frequency: $f_\mathrm{p}\propto n^{1/2} \propto r^{-3/2}$.
An additional indication is an expected similarity of the
velocity--time plot of an eruption to the SXR flux
\citep{Zhang2001, Grechnev2014b, Grechnev2016}. To relate the
spatial and frequency (density) scales, we refer to the top of the
first eruption revealed by the EIT images in
Figure~\ref{F-first_eruption}. The plane-of-sky measurements are
corrected by a factor of $1/\sin{\lambda}$ for the longitude of
AR\,9742 [$\lambda = 54^{\circ}$].

 \begin{figure} 
  \centerline{\includegraphics[width=0.7\textwidth]
   {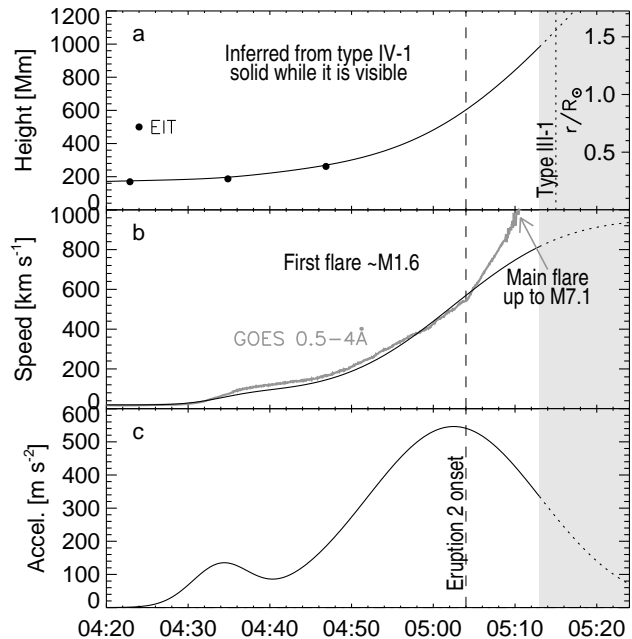}
  }
  \caption{Kinematics of the first eruption (leading edge) in the
radial direction inferred from the analysis of the Type IV-1
burst, SXR flux, and EIT data (filled circles from
Figure~\ref{F-first_eruption}). (a)~Height--time plot.
(b)~Velocity--time plot along with a GOES SXR flux scaled to match
the velocity. (c)~Acceleration--time plot. Type IV-1 is detectable
until the shading. The dashed-vertical line denotes the onset time
of the second eruption. The dotted-vertical line in panel a
denotes the onset time of the Type~III-1 burst.}
  \label{F-rope_plots}
  \end{figure}

We composed the acceleration time-profile from two Gaussian
pulses, adjusting their parameters to make its antiderivative
similar to the SXR flux and to reproduce the Type~IV-1 envelope
(see \citealp{Grechnev2011, Grechnev2013a, Grechnev2014b,
Grechnev2016} for the description of the technique). The
kinematical plots are presented in Figure~\ref{F-rope_plots}. The
calculated low-frequency envelope of Type~IV-1 is shown in Figures
\ref{F-dynamic_spectrum}b and \ref{F-type_IV} by the dashed-green
line. The correspondence of the inferred kinematics to Type~IV-1,
SXR flux, and to the EIT images confirms its likelihood. In spite
of uncertainties, it is clear that when the second eruption
started, the first erupting structure has reached a height of
about 600\,Mm, being still not far away. It stretched closed loops
ahead of the main erupting structure, and cleared the path for its
expansion, thus facilitating its lift-off.

\subsection{Type III Bursts}
 \label{S-type_III}

The \textit{Wind}/WAVES spectrum in
Figure~\ref{F-dynamic_spectrum}d shows three strong Type~III
bursts. Type~III-1 started at the high-frequency edge of the Rad2
passband around 05:15 and lasted within three minutes. Type~III-2
started around 05:27 and had a similar duration. No metric
Type~III bursts are detectable before Type~III-2. Conversely,
Type~III-3 corresponds to a clear group of metric Type IIIs in an
interval of 05:40\,--\,06:00, when minor microwave bursts are
visible at 17\,GHz in Figure~\ref{F-dynamic_spectrum}a. The
sources of these bursts were also located in AR\,9742 (Article~I).

While acceleration of electrons in the flare region is manifested
by the flare emissions in the whole interval of
Figure~\ref{F-dynamic_spectrum}a, the absence of metric Type~IIIs
until Type~III-3 indicates isolation of the magnetic configuration
in AR\,9742 from the interplanetary space. Two minor Type~IIIs in
Figure~\ref{F-dynamic_spectrum}d before Type~III-1 are unlikely to
be important. The configuration opened, when Type~III-3 started.

Type~III-1, with a starting frequency between 14 and 25\,MHz, was
unlikely to have been caused by electrons escaping from the flare
region. The only apparent source of non-thermal electrons is the
expanding flux rope, which appeared in the first eruption and
contained trapped electrons responsible for Type~IV-1.
Reconnection between this flux rope and an open structure such as
a streamer \citep{Grechnev2013a} or coronal hole
\citep{Masson2013} could create a path for electrons trapped in
the flux rope to escape into interplanetary space
\citep{Aschwanden2012}. Besides electrons, protons and heavier
ions, both pre-existing in the flux rope and injected into it in
the course of flare reconnection, from thermal and suprathermal up
to high energies, were released. A rich seed population was
supplied for acceleration by a trailing shock wave. The presence
of the Type~IIIs in the whole frequency range of
\textit{Wind}/WAVES shows that the particles released could reach
the orbit of Earth.

The onset of Type~III-1 (vertical-dotted line in
Figure~\ref{F-rope_plots}a) corresponds to a height of the
flux-rope of $\approx 1.55\,\mathrm{R}_\odot$. Reconnection
between the flux rope and a streamer or coronal hole could occur
at a farthest edge of the flux-rope's flank, at a height of
$\approx 0.78\,\mathrm{R}_\odot$ for a circular geometry. Our
power-law density model does not help here, because the density
multiplier [$n_0$] is unknown. The plasma frequency expected at
this height is 30\,MHz in a streamer \citep{Newkirk1961} and
8.6\,MHz in a coronal hole \citep{Saito1977} \textit{vs.} expected
14\,--\,25\,MHz. Either option is possible with our uncertainties.

While the height\,--\,time plot of the second flux rope formed in
the main eruption is unknown, a similar scenario associated with
Type~III-2 is indicated by Type~IV-2 related to the main flare and
its probable drift to lower frequencies; the starting frequency of
Type~III-2 between 9 and 25\,MHz, similar to that of Type~III-1,
without metric counterparts; and a probable opening of the
magnetic configuration after Type~III-2. The second flux rope
could reconnect with the same open magnetic structure as the first
one and at a comparable height.

The particles accumulated in the second flux rope in the course of a
stronger main flare were released after the Type~III-2 onset. This
population of particles must be more energetic and plentiful. The
estimated SPR time (\citealp{Reames2009b}: the gray diamond labeled
SPR) coincides with Type~III-2. The first particle release during
Type~III-1 does not contradict the GLE onset 12 minutes earlier than
the estimated SPR time (the dashed line in
Figure~\ref{F-high-energy}b); thus, energetic particles accelerated
in the first flare could directly contribute to the SEP event.

Type~III-3 indicates direct escape of electrons (and probably
other particles) from the flare region evidenced by a group of the
corresponding metric Type~IIIs. The peak of the microwave spectrum
shifted from 6\,GHz (Article~II) to $\approx 2.7$\,GHz at that
time, as comparison of the burst at 17\,GHz (black) and 2.7\,GHz
(pink) in Figure~\ref{F-dynamic_spectrum}a shows. This suggests
displacement of the microwave-emitting region (and, possibly, the
site of flare energy release) to weaker magnetic field at larger
altitude. These late-stage processes might be related to the
post-impulsive particle acceleration (see, \textit{e.g.},
\citealp{Chertok1995, Klein1999, Klein2014}). It is possible, but
questionable, that the released particles were accelerated by a
shock wave.

\section{White-Light CME}
 \label{S-CME}

\begin{figure} 
  \centerline{\includegraphics[width=\textwidth]
   {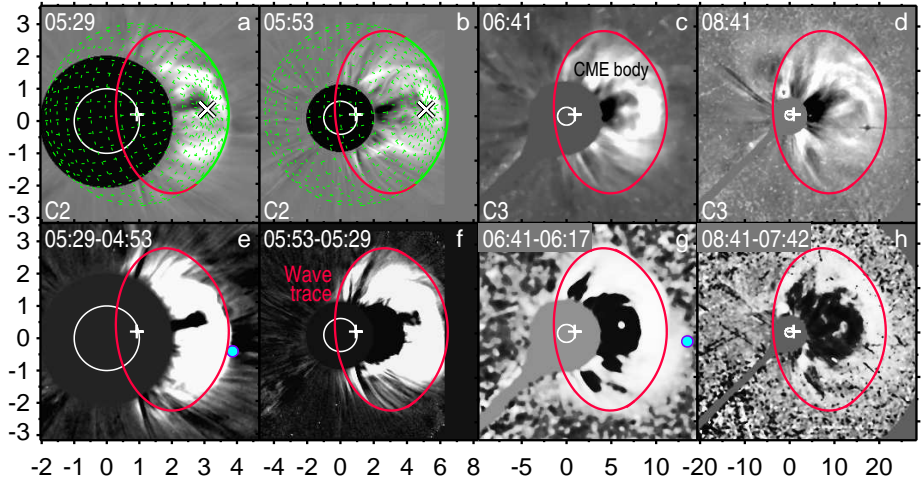}
  }
  \caption{CME body (top) and wave traces (bottom, running differences)
in LASCO-C2 and -C3 images. Panels a and b explain the
calculations of the wave outline (red in other panels). A bulge
protruding southwest is visible in later images. The blue-filled
circles in panels e and g denote the measurements in the CME
catalog corresponding to the bulge. The axes indicate the distance
from solar disk center in solar radii.}
  \label{F-lasco_img}
  \end{figure}

A white-light transient was observed by the LASCO-C2 and -C3
coronagraphs starting from 05:29 up to $30\,\mathrm{R}_\odot$. The
transient consisted of a structured CME body (probable flux rope)
surrounded by a faint partial-halo wave trace. The same LASCO
images in Figure~\ref{F-lasco_img} processed in different ways
reveal CME structures (top) and wave traces (bottom, running
differences), which are detectable from a diffuse halo-like
brightening or deflected coronal rays.

\begin{figure} 
  \centerline{\includegraphics[width=0.6\textwidth]
   {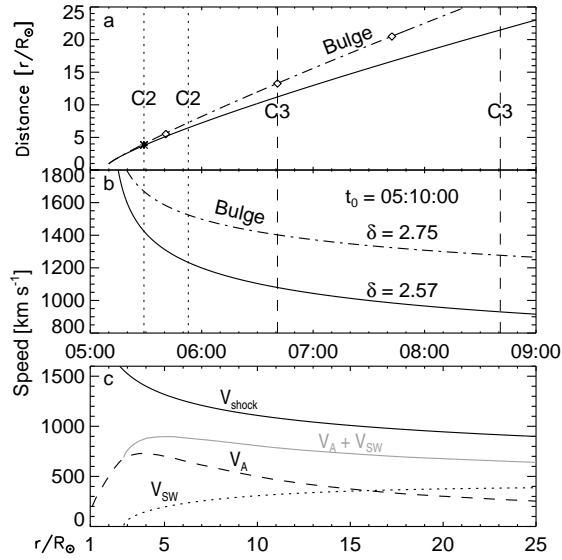}
  }
  \caption{Kinematic plots of the transient visible in LASCO images.
(a)~Height--time plots for the main envelope of the CME leading
edge (solid) and for the bulge (dash--dotted). The symbols
represent the measurements from the CME catalog. (b)~Speed--time
plots for the main CME envelope (solid) and the bulge
(dash--dotted) calculated for a decelerating shock wave with the
same onset time, $t_0 = $\,05:10:00, and different density falloff
exponents $\delta$. (c)~Distance--velocity plot for the main CME
envelope (black solid) corresponding to the wave along with the
models of the Alfv{\' e}n speed $V_\mathrm{A}$ (dashed) and solar
wind speed $V_\mathrm{SW}$ (dotted) above the quiet Sun. The gray
curve represents the sum $V_\mathrm{A} + V_\mathrm{SW}$.}
  \label{F-lasco_plots}
  \end{figure}

The sky-plane expansion of the CME body and wave was very similar.
We therefore fitted the kinematics of both with the same power-law
appropriate for a shock wave. The wave onset time, $t_0
=$~05:10:00, is slightly later than that of shock~2 ($t_{02}
=$~05:09:10). This corresponds to an expected coalescence of two
shock waves into a stronger one with an apparently later onset
time \citep{Grechnev2011}. The power-law exponent is $\delta =
2.57$, close to the mid-latitude Saito model, which describes the
corona above the quiet Sun (different values of $\delta$ found in
Section~\ref{S-type_II} for the Type~II burst were related to a
streamer). The kinematical plots are shown in Figures
\ref{F-lasco_plots}a and \ref{F-lasco_plots}b by the solid lines.
Using these kinematics, the images in Figure~\ref{F-lasco_img} are
progressively resized to maintain the visible size of the
expanding transient (see also the \textsf{2001-12-26\_LASCO.mpg}
movie in the Electronic Supplementary Material.

Figures \ref{F-lasco_img}a and \ref{F-lasco_img}b explain how the
shape of the wave front was calculated. Green is a sphere centered
at the eruption site (straight cross), with a polar axis extending
its radius-vector (slanted cross marks the pole), and a radius
taken from Figure~\ref{F-lasco_plots}a. Red is a small circle on
this sphere (its plane does not pass through the sphere's center).
The red curve in the other panels of Figure~\ref{F-lasco_img},
composed from the red and green arcs in Figures \ref{F-lasco_img}a
and \ref{F-lasco_img}b, matches most wave traces in all images.
Thus, the main part of the wave front is a conic section of the
sphere, as expected for a blast wave. The deflected rays outside
of the outline in Figure~\ref{F-lasco_img}e are most likely due to
the flanks of the earlier shock~1 ($t_{01} =$~05:04). Two shock
waves running across the solar disk one shortly after another were
observed previously \citep{Grechnev2013a}.

Figure~\ref{F-lasco_img}g reveals a southwest bulge ahead of the red
outline, which might be ascribed to a bow-shock, but its orientation
is offset from the main expansion direction of the CME.
Correspondence of the bulge to the position of a large coronal hole
in Figure~\ref{F-eit284} indicates its relation to the fast solar
wind stream from the coronal hole. The measurements in the CME
catalog (\url{cdaw.gsfc.nasa.gov/CME_list/}: \citealp{Yashiro2004})
are related to the bulge. They are denoted by the symbols in
Figure~\ref{F-lasco_plots}a and outlined in Figures
\ref{F-lasco_img}a and \ref{F-lasco_img}b by a power-law fit with a
steeper $\delta = 2.75$ expected for a coronal hole.

Figure~\ref{F-lasco_plots}c presents the velocity \textit{vs.}
distance (black solid) of the main CME envelope related to the
spherical wave front (without the bulge). This is a power-law fit
for the apex of the observed wave front. For comparison, the dashed
curve represents a model for the Alfv{\' e}n speed [$V_\mathrm{A}$]
above the quiet Sun \citep{Mann2003}. The dotted curve shows a model
for the solar-wind speed [$V_\mathrm{SW}$] \citep{Sheeley1997}. If a
wave is weak, then its velocity is $V_\mathrm{SW} + V_\mathrm{fast}$
with $V_\mathrm{fast}$ being the fast-mode speed. The head of the
wave front moves at distances $5-25\,\mathrm{R}_\odot$ practically
along the magnetic field, \textit{i.e.} $V_\mathrm{fast} \approx
V_\mathrm{A}$ ($V_\mathrm{A}$ exceeds the sound speed). The
considerable excess over the Alfv{\' e}n speed of the wave speed
relative to the moving environment $V_\mathrm{shock} > V_\mathrm{A}
+ V_\mathrm{SW}$ (the gray curve shows the sum) confirms the
shock-wave regime in the whole range of distances. The Mach number
characterizing the wave intensity can be estimated as $M =
(V_\mathrm{shock} - V_\mathrm{SW})/V_\mathrm{A}$ (with increasing
$M$ this formula becomes inaccurate). The Mach number is $M \approx
1.6$ at distances $5-10\,\mathrm{R}_\odot$, whereas $M \approx 2$ at
$25\,\mathrm{R}_\odot$.

In summary, the partial halo surrounding the CME body was most
likely a trace of a shock wave, which had both blast-wave and
bow-shock properties in the LASCO field of view from
$3.8\,\mathrm{R}_\odot$ to $30\,\mathrm{R}_\odot$. The CME and
wave were super-Alfv{\' e}nic and expanded similarly (bow-shock),
while their common kinematics and the spherical wave front
corresponded to the impulsively excited blast wave, whose
propagation was controlled by the growing mass of the swept-up
plasma.

\section{Discussion}
 \label{S-discussion}

\subsection{Presumable Kinematics of the Fast CME}

The results of the analysis allow reconstruction of the presumable
kinematics of the fast CME. We also invoke the following
conclusions and observational results: i)~acceleration of a CME is
synchronous with an HXR (or microwave) burst \citep{Zhang2001,
Temmer2008, Temmer2010, Grechnev2015b}; ii)~the velocity--time
plot of a CME is similar to the SXR flux \citep{Zhang2001,
Grechnev2016}; iii)~the Neupert effect: the temporal profile of an
HXR (microwave) burst is similar to the derivative of the SXR flux
\citep{Neupert1968}; and iv)~the height--time and velocity--time
plots of the CME in the LASCO field of view should be close to
those of the shock wave in Figures \ref{F-lasco_plots}a and
\ref{F-lasco_plots}b (Section~\ref{S-CME}).

\begin{figure} 
  \centerline{\includegraphics[width=0.75\textwidth]
   {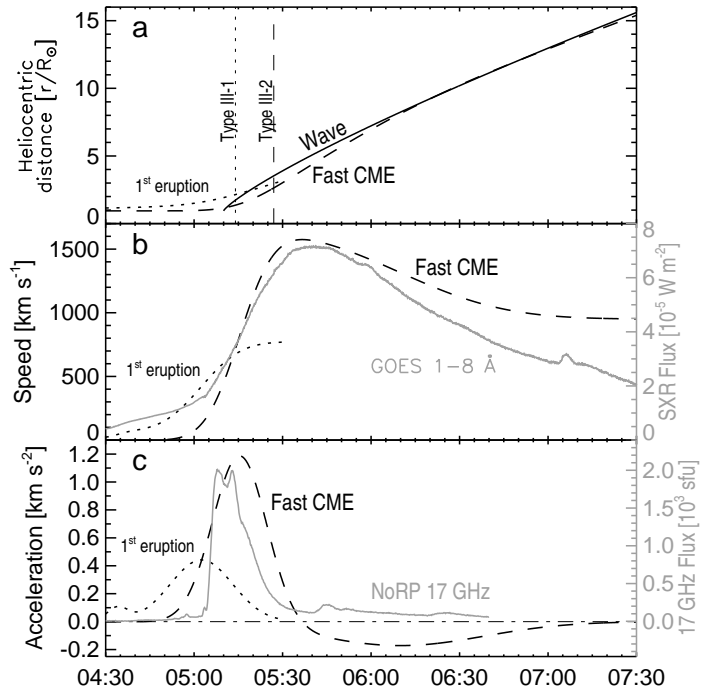}
  }
  \caption{Presumable plane-of-the-sky kinematics of the fast CME
(dashed) up to $15\,\mathrm{R}_\odot$ inferred from indirect
indications. (a)~Distance--time plot and a wave plot from
Figure~\ref{F-lasco_plots}a (solid). (b)~Velocity--time plot along
with the GOES SXR flux (gray). (c)~Acceleration--time plot and a
microwave burst at 17\,GHz (gray). The dotted curves correspond to
the first eruption (Figure~\ref{F-rope_plots}) inferred from the
analysis of Type~IV-1 and corrected to the plane of the sky. The
broken vertical lines in panel a denote the onset times of the
Type~III-1 and Type~III-2 bursts.}
  \label{F-presumable_plots}
  \end{figure}

According to the last item, the initial impulsive acceleration of
the fast CME should be followed by a deceleration phase, because
the shock wave decelerated all of the time. We therefore composed
the CME acceleration as a positive Gaussian pulse resembling the
microwave burst followed by a negative pulse. The velocity should
be roughly similar to the GOES 1\,--\,8\,\AA\ flux, being about
1000\,km\,s$^{-1}$ at 07:30. The result is shown in
Figure~\ref{F-presumable_plots} by the dashed curves. The dotted
curves correspond to the first eruption (Figure~\ref{F-rope_plots}
in Section~\ref{S-type_IV}) corrected to the plane of the sky by a
factor of $\sin{\lambda}$ (with a longitude $\lambda = 54^\circ$).
The CME and shock wave (solid lines) in
Figure~\ref{F-presumable_plots}a are close to each other after
05:30, as expected.

The presence of two eruptions forming the CME complicates the
situation. The flux rope ejected during the first flare eventually
constituted the CME top part and the second flux rope joined it
from below. The difficulties in untangling the kinematical plots
of the two CME components result in the differences between the
CME speed and SXR flux in Figure~\ref{F-presumable_plots}b and
between the CME acceleration and microwave flux in
Figure~\ref{F-presumable_plots}c. The derivative of the SXR flux,
which contains an increasing component from the first flare, is
smoother than the 17\,GHz burst. We have adopted a compromise
shape of the acceleration somewhat smoother than the microwave
burst and a corresponding velocity somewhat sharper than the SXR
flux. The estimated acceleration peak (1.2\,km\,s$^{-2}$ in the
plane of the sky and 1.5\,km\,s$^{-2}$ in the radial direction) is
comparable with a radial acceleration of 1.1\,km\,s$^{-2}$
estimated  by \cite{Gopalswamy2012} for this event from different
considerations. Our value might still be underestimated;
nevertheless, the plots appear to be acceptable with the
uncertainties we have.

The CME plots in Figure~\ref{F-presumable_plots} were
reconstructed in the basis of recent results by referring to
dynamic radio spectra and scarce imaging observations. These plots
lead to the following conclusions discussed in the next sections.

\begin{enumerate}

\item The acceleration of second flux rope corresponds to a
typical impulsive piston. It must have excited a blast-wave-like
shock not later than the acceleration peak. The second flux rope
was below the first one at that time.

\item The second flux rope exceeded the first one in speed and
acceleration by a factor of $\geq 2$ and became super-Alfv{\'
e}nic before its observations by LASCO.

\item The heights reached by the first flux rope and the second
one, when corresponding Type~III bursts started, are comparable,
being related as 1:1.35.

\end{enumerate}

\subsection{Shock Wave}

\subsubsection{Shock-Wave History}

The observations found for this event fit within a scenario
outlined in Section~\ref{S-introduction}. The second flux rope
formed by reconnection (also responsible for the flare)
accelerated and produced a strong disturbance as an impulsive
piston. The shock-wave excitation by the jet-like eruption was
similar. The impulsive-piston scenario is well known -- see,
\textit{e.g.}, \cite{VrsnakCliver2008}. A crucial factor for the
shock formation not considered in this article is inhomogeneous
distribution of the fast-mode speed. Propagating into environment
of a much lower fast-mode speed, the disturbance undergoes jamming
of its profile and must rapidly steepen into the shock
\citep{Afanasyev2013}.

The impulsive-piston excitation of a shock wave and its properties
resemble the expectations for the hypothetical ignition of a freely
propagating decelerating blast wave by the flare pressure pulse.
However, the role of the impulsive piston in the observed events is
played by the erupting flux rope rather than the flare loops. The
flare-ignition of shock waves is neither supported by observations
nor expected from general considerations for the following reasons.

The pressure of the plasma in flare loops is controlled by its
temperature and density, which determine the SXR emission of the
loops. It is intrinsically gradual and usually resembles the
antiderivative of the HXR or microwave burst (the Neupert effect:
\citealp{Neupert1968}). On the other hand, the HXR burst is
similar to the acceleration pulse of an erupting structure, as
several studies concluded (see Section~\ref{S-introduction}).
Thus, any erupting structure is generally an efficient impulsive
piston, producing much sharper pressure pulse than the flare
loops.

Even the situation with the plasma $\beta$ in flare loops $\beta
\approx 1$ is normal in a flare. The plasma-pressure increase
caused by chromospheric evaporation is balanced by the dynamic
pressure of the reconnection outflow. All dimensions of a flare
loop increase by an inconsiderable factor of $\sqrt[4]{1+\beta}$
\citep{Grechnev2006beta}. The disturbance is too weak to produce a
shock wave. Our observational studies showed that the size of
SXR-emitting flare loops does not change when a shock wave
appears, and that its onset time is close to the acceleration peak
of an erupting structure, which can precede the HXR peak by one to
two minutes.

We found that shock waves were impulsively excited by sharply
erupting flux ropes in eight events, when their kinematics were
measured. These events range from the GOES B- to X-class and were
or were not accompanied by non-thermal HXR and/or microwave
bursts. The shock waves in these events and three others, in which
their exciters were not measured, exhibited identical behaviors,
initially resembling blast waves. In the events with fast CMEs,
shock waves showed some properties of bow shocks only after some
time. The pure bow-shock excitation scenario has not been found in
any flare-related eruption. The bow-shock regime presumed in most
studies of SEPs (\citealp{Reames2009a, Aschwanden2012,
Gopalswamy2012}; and others) is ruled out by the fact that two
shock waves observed in two events followed each other within six
minutes. The initial bow-shock excitation is not excluded for
gradually accelerating non-flare-related CMEs.

The 26 December 2001 event associated with a major flare shows the
same shock-wave history as the flare-related events studied
previously. The first shock wave was impulsively excited by the
main eruption at 05:04. The second shock was similarly excited by
the jet at 05:09. Both shock waves following each other initially
resembled decelerating blast waves. The trailing front must have
reached the leading one around the radial direction and merged
with it into a stronger one with a later onset time at 05:10. As
Section~\ref{S-CME} concluded, the partial halo surrounding the
CME body was a trace of a shock wave, whose properties in the
LASCO field of view were intermediate between the blast wave and
bow shock.

\subsubsection{Regimes of Shock Waves}

The solid curve in Figure~\ref{F-lasco_plots}c is a power-law fit
of the measured shock-wave speed and formally corresponds to a
solution of a self-similar blast wave in plasma with a power-law
density \textit{vs.} distance dependence. The mass of plasma
involved in the motion continuously grows. The integral of the
kinetic energy is conserved, as well as the integral of the sum of
the plasma thermal energy and magnetic-field energy within the
volume behind the shock front. Conservation of both energy
integrals in this regime suggests an increasing impulse of the
mass moving within a fixed solid angle, if the density falls off
no more steeply than $r^{-3}$ (to have a finite initial mass). The
impulse increases because of a considerable pressure difference
upstream and downstream of the shock front. The system consisting
of the CME and associated shock possesses such properties. The
initial impulse of the system is zero, the total mass of the
moving gas grows, and the impulse increases due to magnetic
driving forces responsible for the development and expansion of
the CME. This situation persists up to some distance. Our
power-law fit applies to the position and speed of the
decelerating shock front at this stage.

The magnetic driving forces, which initially increased the CME
impulse, weaken with an increasing distance from the Sun. The
outer magnetic influence on the CME and disturbed solar wind
ceases. The total impulse of the system starts being conserved.
The CME deceleration is governed by the interaction regime of the
CME with the solar wind. Two extremities are possible: i)~flow
around the CME without change in its mass and ii)~the ``snowplow''
regime, when the mass of the CME moving by inertia grows because
of adhesion of the solar wind plasma to the CME instead of flowing
around it. The change in the CME impulse in either regime is equal
(but opposite) to that of the solar wind. The total drag force is
equal to the time-derivative of this impulse and can be
calculated. We do not consider the ``snowplow'' regime, which
results in a much stronger deceleration than the solid curve in
Figure~\ref{F-lasco_plots}c shows. The mass increase is unlikely
at large distances from the Sun, where the CME expands radially.

Two situations of the flow around the CME are possible: i)~the
solar-wind flow rapidly recovers behind a compact CME and ii)~the
disturbed zone behind the CME is much larger than that ahead of
it. The drag force is absent in the idealized first situation,
being largest in the second. If the difference between the
velocities of the CME leading edge and solar wind [$V_\mathrm{CME}
- V_\mathrm{SW}$] exceeds the fast-mode speed ahead the CME (as
was the case in our event), then bow shock exists ahead of it with
a speed practically equal to $V_\mathrm{CME}$.

A model of the stationary solar wind is usually invoked to
calculate $V_\mathrm{fast}$. In a real situation, the CME
formation is associated with the appearance of a large-scale
fast-mode shock wave. Propagating up like a blast wave, it leaves
behind an extended region, where $V_\mathrm{fast}$ is higher than
the calculated value, so that $V_\mathrm{CME} - V_\mathrm{SW} <
V_\mathrm{fast}$ for some time. The CME expansion at this stage
has the character of a flow and the bow shock is absent.

After that, a bow shock can appear even during the early CME
expansion, but its role is insignificant, as long as the driving
forces surpass the drag force. Only after the driving forces
diminish relative to the drag force does the latter dominate the
CME kinematics. Deceleration of both a freely propagating blast
wave and bow shock ahead of a fast CME influenced by the
aerodynamic drag complicates identification of the shock-wave
regime. To assess the role of the drag at different distances, we
will compare the observed velocity--distance plot with
expectations for the bow-shock regime. We find the
$V_\mathrm{CME}(t)$ dependence required to calculate the bow-shock
curves from the motion equation
\begin{equation}
 m_\mathrm{CME}(\mathrm{d}V_\mathrm{CME}/\mathrm{d}t)=-\rho(V
_\mathrm{CME}-V_\mathrm{SW})^{2}S \equiv F_\mathrm{drag},
 \label{E-motion}
 \end{equation}
\noindent where $m_\mathrm{CME}$ is a constant effective mass of
the CME and $S$ is its cross-section. At a sufficient distance
$S\propto r^{2}$. For simplicity we take $V_\mathrm{SW}$ at a
reference distance $r_\mathrm{ref}$, neglecting its gradual
variation in Equation~(\ref{E-motion}). Due to conservation of the
plasma flow in a stationary solar wind with a density $\rho$,
$\rho S=\rho_\mathrm{ref} S_\mathrm{ref}$ is constant. The
``$\mathrm{ref}$'' subscript corresponds to a distance
$r_\mathrm{ref}$ or time $t_\mathrm{ref}$. We transform
Equation~(\ref{E-motion}) to the form
\begin{equation}
\mathrm{d}V_\mathrm{CME}/\mathrm{d}t = -C(V
_\mathrm{CME}-V_\mathrm{SW})^{2},
 \label{E-motion-2}
\end{equation}
\noindent with $C=\rho_\mathrm{ref}
S_\mathrm{ref}/m_\mathrm{CME}$. The solution of
Equation~(\ref{E-motion-2}) is a function
\begin{equation}
V_\mathrm{CME} = V_{\mathrm{CME}} (t_\mathrm{ref})
\left(1+\kappa_{\mathrm{ref}}\frac{t-t_\mathrm{ref}}{\tau_\mathrm{ref}}\right)
\left(1+\frac{t-t_\mathrm{ref}}{\tau_\mathrm{ref}}\right)^{-1}
 \label{E-V_CME}
\end{equation}
 \noindent with $\kappa_{\mathrm{ref}} = V_\mathrm{SW}/
V_{\mathrm{CME}}(t_\mathrm{ref})$,
$\tau_\mathrm{ref}=[C(V_{\mathrm{CME}}(t_\mathrm{ref})-
V_\mathrm{SW})]^{-1} = -[V_{\mathrm{CME}}(t_\mathrm{ref})-
V_\mathrm{SW}]/a_\mathrm{ref}$ being deceleration time scale, and
$a_\mathrm{ref}=(F_\mathrm{drag}/M)_{t=t_\mathrm{ref}}= $ \\
$[\mathrm{d}V_\mathrm{CME}/\mathrm{d}t]_{t=t_\mathrm{ref}}$ being
the acceleration at a reference point. The characteristic
deceleration time [$\tau_\mathrm{ref}$] is obtained by
differentiation of Equation~(\ref{E-V_CME}):
\begin{equation}
\tau_\mathrm{ref}=
(\kappa_{\mathrm{ref}}-1)V_{\mathrm{CME}}(t_\mathrm{ref}) /
[{\mathrm{d}V_\mathrm{CME}}/{\mathrm{d}t}]_{t=t_\mathrm{ref}}.
 \label{E-tau}
\end{equation}

\begin{figure} 
  \centerline{\includegraphics[width=0.65\textwidth]
   {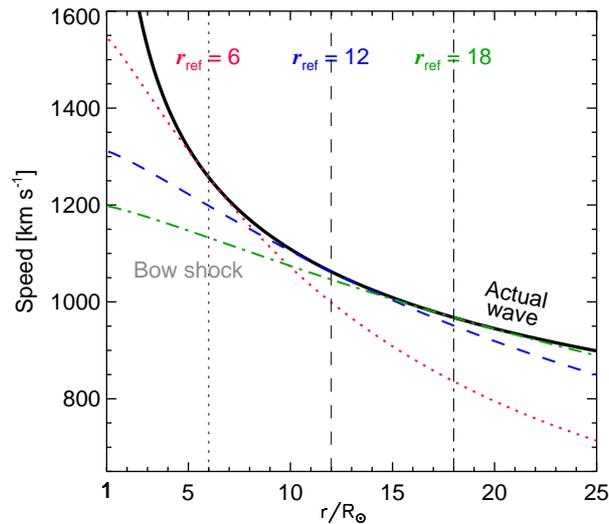}
  }
  \caption{Comparison of the actual distance--speed wave evolution
(same as in Figure~\ref{F-lasco_plots}c) with the plots expected
for the bow-shock regime with different parameters (colored
curves). The broken-vertical lines mark the reference distances
$r_{\mathrm{ref}}$, at which the speed and deceleration of the
actual wave were taken to calculate corresponding plots for the
bow-shock regime (denoted by the same colors and line styles).}
  \label{F-bow_shock}
  \end{figure}

Figure~\ref{F-bow_shock} illustrates how the changing relation
between the driving forces and drag force affects the CME
deceleration. The black-solid line is a power-law fit of the
observed velocity \textit{vs.} distance dependence for the shock
wave. The broken colored curves represent the trajectories [$V
_\mathrm{CME}(r)$] calculated from Equations (\ref{E-V_CME}) and
(\ref{E-tau}) for a constant CME mass. Each curve for the
bow-shock regime is calculated by referring to the speeds of the
shock and solar wind at three reference distances
$r_\mathrm{ref}=[6, 12, 18]\,\mathrm{R}_\odot$. Drag is formally
assumed to dominate at points $r_\mathrm{ref}$, and the speeds of
the shock front and CME are equal, $V_\mathrm{shock}= V
_\mathrm{CME}$, \textit{i.e.} the wave is bow shock. If the drag
force surpassed the driving forces in the whole range of
distances, then all four curves in Figure~\ref{F-bow_shock} would
have coincided. In reality, the calculated curve approaches the
experimental plot only at distances $r\geq r_\mathrm{ref} \approx
15\,\mathrm{R}_\odot$, where aerodynamic drag can determine the
CME deceleration.

\subsubsection{Interplanetary Type II Event}

We established consistency of a predicted Type~II trajectory with
observations up to a very low frequency of 250\,kHz
(Figure~\ref{F-type_II_Cane}). This result reconciles the Type~II
emissions in the metric range with longer waves and addresses a
long-standing discussion over their seemingly different origins
(\textit{e.g.} \citealp{CaneErickson2005}). Some other properties of
the IP Type~IIs become clearer.

A narrow-band Type~II burst can only appear from a compact source
in a narrow structure such as a coronal ray \citep{Uralova1994,
Reiner2003}. Otherwise, a large shock front crossing a wide range
of plasma densities produces a drifting continuum
\citep{KnockCairns2005}. Coexisting signatures of two shocks in
the Type~II emission until 09:00 (Section~\ref{S-type_II}) suggest
a location of, at least, one of the sources at a flank of the
shock wave. On the way to an observer, the emission crossing a
dense heliospheric plasma sheet may be subjected to refraction,
interference, and/or absorption. The IP Type~II emission from a
moving compact source can fade and reappear, producing ``blobs and
bands'' structures \citep{CaneErickson2005}. The structures of IP
Type~IIs observed from the vantage points of \textit{Wind} and
STEREO can therefore be different.

\subsection{Particle Release}

\subsubsection{Type III Bursts and Particle Release}

While remote-sensing methods for protons and heavier ions are
limited by detection of $\gamma$-rays emitted in their
interactions with dense material (\textit{e.g.}
\citealp{Vilmer2011}), electrons can be used as their probable
tracers. Their signatures extend from $\gamma$-ray bremsstrahlung
continuum up to long radio waves. In the course of an eruptive
flare, electrons and heavier particles are presumably injected
from the reconnection site both down, into the flare loops, and
up, into the forming flux rope. Electrons confined in a rising
flux rope (which can also contain heavy particles) may produce a
drifting Type~IV burst. If reconnection between the flux rope and
an open structure grants the trapped particles access to the
interplanetary space, then a DH Type~III burst produced by
escaping electrons hints at a possible release of heavier
particles.

Section~\ref{S-type_III} considered release of electrons trapped
in the first flux rope after reconnection of its flank with a
streamer or coronal hole producing Type~III-1. Its onset
corresponds to heights of the flux-rope's top of
$1.55\,\mathrm{R}_\odot$ and the reconnection site of
$0.78\,\mathrm{R}_\odot$. \cite{Masson2013} showed a possibility
of SEP release in reconnection between a flux rope and coronal
hole. In our event, the distance between AR\,9742 and the southern
coronal hole in Figure~\ref{F-eit284} projected to this height was
about $1\,\mathrm{R}_\odot$. The extra-radial divergence of
magnetic field lines in a coronal hole makes its contact with a
flux rope at this height possible. The left-handed flux rope with
an axial field pointed the Northeast can reconnect with the
southeasterly S-polarity coronal hole.

The appearance of a path to the interplanetary space could also
release heavier particles trapped in the first flux rope before
the Type~III-1. Appropriate species supplied a rich seed
population for acceleration by a trailing shock wave, which
arrived in about five minutes, as Figure~\ref{F-presumable_plots}a
shows. This probably occurred at one of the flanks of the flux
rope and shock wave rather than at their noses.

The estimated height of the second flux-rope's top at the onset of
Type~III-2 is $2.1\,\mathrm{R}_\odot$. The height of the
presumable reconnection site is $\approx 1.05\,\mathrm{R}_\odot$.
Reconnection between the second flux rope expanding extra-radially
and the coronal hole at this greater height is also probable. The
starting frequency of Type~III-2 expected at this height is
6.6\,MHz (or 21\,MHz for reconnection with the streamer); actually
it was absent above 25\,MHz and certainly present below 9\,MHz, as
Figure~\ref{F-dynamic_spectrum} in Section~\ref{S-type_III} shows.
Its onset time, 05:27, is close to the SPR time of
05:29:00\,$\pm3.7$\,minutes estimated by \cite{Reames2009a,
Reames2009b}. However, the heliocentric distances of the particle
release during Type~III-1 and Type~III-2 that we estimate for a
CME flank are roughly about $2\,\mathrm{R}_\odot$ \textit{vs.}
$(3.6 \pm 0.5)\,\mathrm{R}_\odot$ estimated by \cite{Reames2009a,
Reames2009b} for the CME nose.

The absence of metric Type~III bursts during DH Type~III-1 and
Type~III-2 under the presence of accelerated electrons evidenced
by flare emissions (Figure~\ref{F-dynamic_spectrum}) indicates
confinement of the magnetic configuration at that time. In
contrast, the appearance of a dense metric Type~III group
associated with DH Type~III-3 indicates that the configuration
opened. Particles accelerated in the flare region gained direct
access to the interplanetary space. Thus, the Type~III-3 indicates
a post-impulsive particle acceleration argued for in several
studies (\textit{e.g.} \citealp{Chertok1995, Klein1999,
Klein2014}). Although the related particle flux seems to be weaker
than in the main flare, its contribution can be appreciable at
moderate energies due to the long durations of post-eruption
processes.

\subsubsection{Particle Release in Other GLE Events}

The preceding section confirms that DH Type~III bursts
(1\,--\,14\,MHz) can trace release of heavy particles. To test how
common this pattern is, we examined the GLEs analyzed by
\cite{Reames2009a}, for which \textit{Wind}/WAVES data are
available. These are 13 out of 16 GLEs of Solar Cycle 23,
excluding GLE55, GLE62, and GLE68, whose analysis was problematic.
From one to four strong DH Type~III bursts were observed in each
event. The SPR time estimated by \cite{Reames2009b} from the VDA
analysis is typically close to one of these DH Type~IIIs and never
precedes their group. Coincidence is present in seven events, the
interval between a Type~III burst and the estimated SPR error band
is within five minutes in four events, and the SPR time is
considerably later than the Type~IIIs in GLE58 (24 August 1998,
$\approx 30$\,minutes) and GLE66 (29 October 2003, $\approx
15$\,minutes).

Following the traditional hypotheses, \cite{Reames2009a,
Reames2009b} assumes that the shock wave starts to form when the
CME exceeds the Alfv{\' e}n speed ahead of it. The onset of a
Type~II burst is regarded as the onset of the shock formation. The
shock is assumed to strengthen afterwards. The SPR time is
considered as a sign when the particle acceleration becomes
efficient. However, the preceding section indicates that shock
waves appear during the flare rise, being able to accelerate
particles much earlier than traditionally assumed.

\cite{Reames2009b} found for the 29 October 2003 event (GLE66) an
SPR time of 20:55.6\,ST\,$\pm 5.8$\,minutes, \textit{i.e.}
21:03:56\,UTC\,$\pm 5.8$\,minutes for an observer on Earth.
According to \cite{Balasub2007}, a Moreton wave in this event
started as early as 20:43\,UTC and propagated with a speed of
1100\,--\,1200\,km\,s$^{-1}$, which establishes its shock-wave
regime. A Type~II burst was observed at Culgoora (consistent with
the Palehua spectrogram) starting at 430\,MHz from 20:42\,UTC,
close to the Moreton wave onset. GLE66 started at 21:05\,UTC
\citep{Gopalswamy2012}, and the onset of the $> 100$\,MeV proton
enhancement is detectable in the GOES-10 data at the same time.
With a magnetic path length of $1.75 \pm 0.09$\,AU
\citep{Reames2009b}, the estimated SPR time seems to be too late.

Note that GLE66 occurred during the strongest Forbush decrease
since the 1980s. It was caused by a huge magnetic cloud moving
with a very high speed of 1900\,km\,s$^{-1}$
\citep{Grechnev2014a}. Its rapid motion affected magnetic path
lengths for different-energy particles, distorting the results of
VDA.

Another exception is GLE58 on 24 August 1998. \cite{Reames2009b}
measured the SPR time for this event at 22:32.1\,ST\,$\pm
4.6$\,minutes, \textit{i.e.} 22:40:26\,UTC\,$\pm 4.6$ minutes.
According to \cite{Vrsnak2002}, a Moreton wave started at 22:03\,UTC
with an initial speed of 946\,km\,s$^{-1}$, decelerated like a blast
wave, and covered 500\,Mm. A Type II burst started at 22:02\,UTC,
close to the Moreton wave onset. The shock wave started early in
this event. GLE58 started at 22:50\,UTC \citep{Gopalswamy2012}. A
proton enhancement $> 100$\,MeV is detectable at that time. With an
estimated path length of $1.55 \pm 0.04$\,AU, the 100\,MeV protons
should have left the Sun by 22:29\,UTC, well before the estimated
SPR time, which is also challenging to understand.

There was no conspicuous anomaly in the Earth's environment during
the GLE58 occurrence. A key can also be related to
energy-dependent transport effects (drifts, diffusion, and
others), whose importance is prompted by a moderately eastern
position of the solar source region (E09\,N35). Impressive
transport issues were demonstrated by the 1 September 2014 solar
event behind the east limb. The rise phase of a related proton
enhancement was dominated during half a day by $>100$\,MeV
protons, whereas their expected flux should be suppressed.

Most likely, the uncertainties of the VDA results obtained by
\cite{Reames2009b} for the 13 GLEs of Solar Cycle 23 were
underestimated because of transport effects not considered
(\citealp{Laitinen2015} analyzed some corrections in the VDA).
Presumably realistic SPR times in these events were close to the DH
Type~III bursts, suggesting concurrent release of heavy particles
and electrons, probably accelerated by flare processes. A detailed
analysis of the 2 May 1998 (GLE56) and 2 November 2003 (GLE67)
events by \cite{Kocharov2017} has led to a similar scenario.

However, the VDA has not revealed an expected earlier population
accelerated by shock waves, which should appear during the flare
impulsive phase. The shock-accelerated population should exist,
dominating at low energies. Possibly, the seed population is
supplied by a preceding eruption, as in our event. In this case,
ions released from a flux rope should inherit the properties of a
flare region such as a high iron ionization state (\textit{cf.}
\citealp{Desai2006}), but this is not always the case (see
Section~\ref{S-SEP}). Possibly, material of quiet coronal
structures swept up by the top of an expanding flux rope can be
implicated. These circumstances call for rethinking the conditions
of particle acceleration by shock waves and the signatures of
these particles.

Note that magnetic confinement of protons and heavier ions is not
as tight as that of electrons. Having much larger gyroradii, heavy
particles can escape more easily than electrons. Therefore, DH
Type~III bursts hint at the most effective release of heavy
particles, while their escape is possible at different times.

The flux-rope-mediated escape of accelerated protons and electrons
is different. The mean free path of a fast ion with a mass
[$m_{\mathrm i}$], charge [$e_{\mathrm i}$], and an initial velocity
[$v_0$] in plasma with a number density [$n$] is $\lambda_{\mathrm
i} = m_{\mathrm i} m_{\mathrm e} v{_0}{^4}/ \left( 16 \pi e_{\mathrm
i}^2 e_{\mathrm e}^2 \Lambda n\right)$, where $\Lambda \approx 10$
is Coulomb logarithm, [$e_{\mathrm e}$] and [$m_{\mathrm e}$] are
the electron charge and mass; $\tau_{\mathrm{coll}} \approx
\lambda_{\mathrm i}/v_0$. The lifetime of relativistic electrons
with an energy $E_\mathrm{e}$ is $\tau_{\mathrm{life\,(e)}} \approx
2.6 \times 10^9 E_\mathrm{e}/n$. For example, the lifetime of
100\,MeV protons exceeds the lifetime of 0.5\,MeV electrons in the
same plasma by two orders of magnitude. Thus, the electron-to-proton
ratio escaping from a flux rope with an initial $n >
10^{10}$\,cm$^{-3}$ may be much less than for their direct escape
from the flare site.

\section{Summary}
 \label{S-summary}

Combining observations of the 26 December 2001 event in various
spectral ranges, we reconstructed its scenario and histories of
the CME and shock wave. This solar event consisted of the
following episodes. i)~The first flux-rope's eruption started
around 04:30. ii)~The second, largest eruption around 05:04
produced the first shock wave and the main CME. iii)~The third,
jet-like eruption around 05:09 produced the second shock wave.
Each wave was most likely impulsively excited by an abruptly
erupting flux rope, rapidly steepened into a shock due to a steep
falloff of the fast-mode speed outward from the eruption region,
and initially resembled a blast wave. Both waves ultimately merged
around the radial direction into a single stronger shock wave.
Being followed by a fast CME body, it should eventually change to
the bow-shock regime.

\begin{enumerate}

 \item
The shock wave within the LASCO field of view was in an
intermediate regime between the two extremities of blast wave and
bow shock. The wave kinematics was controlled by the trailing CME,
whose mass grew at the first stage because of the swept-up plasma.
This factor, missed in many studies, determined strong
deceleration of the wave at this stage, different from the
bow-shock regime, which becomes possible at distances $>
15\,\mathrm{R}_\odot$.

\quad The Type~II emission in this event was traced from meters up
to $\approx 250$\,kHz. Two shock waves coexisted at moderate
distances from the Sun. This rules out the bow-shock regime at
this stage and indicates location of, at least, one of the radio
sources at a flank of the shock. Correspondence of a calculated
trajectory to the overall observed evolution of the Type~II
emissions in the whole frequency range demonstrates their common
origin.

\quad The shock excitation scenario described here was the only
one in eruptive flares that we studied so far with a GOES
importance from B to X class. Neither did we observe a different
overall history of the shock wave ahead of a fast CME.

 \item
Shock waves appear during the flare rise, being able to accelerate
particles much earlier than usually assumed. A delayed particle
release time suggests instead their acceleration during the flare
and accumulation in the flux-rope's magnetic trap until the access
to the interplanetary space appears, which is possible in
reconnection of the flux rope with an open structure. The rate of
particle escape from the trap (\textit{i.e.} their flux) can
considerably exceed the rate of their injection into the trap
during the flare. This transport scenario can account for the
contrast between the strong proton flux and a microwave burst,
which was not extreme on 26 December 2001.

\quad The flux-rope-mediated transport scenario is supported by
the closeness of the estimated particle release to the DH Type~III
bursts in most GLE events of Solar Cycle 23. This scenario can
also supply the seed population for acceleration by a trailing
shock wave. In this case, particles are most likely released at
the flank of an expanding flux rope and shock wave. This results
in considerably lesser heights of particle release than usually
assumed.

 \item
The first eruption stretched closed structures above the active
region, facilitating escape of flare-accelerated particles and
lift-off of the main CME. Having not spent a part of its energy to
overcome the magnetic tension of closed structures and to sweep up
plasma ahead, the main CME could reach a higher speed and drive a
stronger shock. Thus, the preceding eruption could have amplified
the outcome of both flare-accelerated and shock-accelerated
protons. Another factor was excitation of two shock waves,
eventually merging in a stronger shock. Both of these factors can
amplify solar particle events.

\end{enumerate}

\begin{acks}

We thank B.I.~Lubyshev, N.V.~Nitta, I.M.~Chertok, and H.V.~Cane for
discussions and assistance, and Y.~Kubo for preparing the HiRAS
spectra for us. We are indebted to an anonymous reviewer for
valuable remarks. We thank the instrument teams of EIT, LASCO, and
MDI on SOHO (ESA and NASA), \textit{Wind}/WAVES, GOES, TRACE, and
MLSO operated by NCAR/HAO; NICT (Japan); USAF RSTN Network; and the
CME Catalog at the CDAW Data Center (NASA and Catholic University of
America) for the data used in this study. The study was supported by
the Russian State Contracts No.~II.16.3.2 and No.~II.16.1.6.
A.~Kochanov was supported by the Russian Foundation of Basic
Research under grants 15-02-03717 and 15-02-01089. V.~Kiselev was
supported by the Marie Curie PIRSES-GA-2011-295272 RadioSun project.

\end{acks}

\section*{Disclosure of Potential Conflicts of Interest} The authors
claim that they have no conflicts of interest.

\end{article}

\end{document}